\documentclass[a4paper,twocolumn,11pt, accepted=2026-01-19]{quantumarticle}
\pdfoutput=1
\usepackage[utf8]{inputenc}
\usepackage{algorithm}
\usepackage{algpseudocode}
\usepackage{blindtext}
\usepackage{bm}
\usepackage{braket}
\usepackage{hyperref}
\usepackage{amssymb}
\usepackage{amsmath}
\usepackage{graphicx}
\usepackage{float}
\usepackage{bbold}
\usepackage{xcolor}
\usepackage{mathtools}
\usepackage{color,tikz,float}
\usepackage{dsfont}
\usepackage{tikz-network}
\usepackage{enumitem}
\usepackage{mathrsfs}
\usepackage{stmaryrd}
\usepackage{multirow}
\usepackage{mathabx}
\usepackage[normalem]{ulem}

\DeclareRobustCommand{\bbone}{\text{\usefont{U}{bbold}{m}{n}1}}

\usepackage[numbers,sort&compress]{natbib}

\usepackage{listings}
\definecolor{codegreen}{rgb}{0,0.6,0}
\definecolor{codegray}{rgb}{0.5,0.5,0.5}
\definecolor{codepurple}{rgb}{0.58,0,0.82}
\definecolor{backcolour}{rgb}{0.95,0.95,0.92}
\lstdefinestyle{mystyle}{
    backgroundcolor=\color{white},   
    commentstyle=\color{codegreen},
    keywordstyle=\color{magenta},
    numberstyle=\tiny\color{codegray},
    stringstyle=\color{codepurple},
    basicstyle=\ttfamily\footnotesize,
    breakatwhitespace=false,         
    breaklines=true,                 
    captionpos=b,                    
    keepspaces=true,                 
    numbers=left,                    
    numbersep=5pt,                  
    showspaces=false,                
    showstringspaces=false,
    showtabs=false,                  
    tabsize=2
}
\algrenewcommand\algorithmicrequire{\textbf{Input:}}
\algrenewcommand\algorithmicensure{\textbf{Output:}}
\newcommand{\I}{\mathrm{i}}

\renewcommand{\L}[1]{\mathcal{L}\left( #1 \right)}
\newcommand{\Tr}{\mathrm{Tr}}
\newcommand{\mc}[1]{\mathcal{#1}}
\newcommand{\mr}[1]{\mathrm{#1}}
\newcommand{\mf}[1]{\mathfrak{#1}}
\newcommand{\mb}[1]{\mathbb{#1}}
\renewcommand{\t}[1]{\textrm{#1}}

\newtheorem{theorem}{Theorem}[section]

\newtheorem{fact}[theorem]{Fact}

\usepackage{hyperref}
\usepackage[printonlyused,withpage]{acronym}

\begin{document}
\title{QMetro++ - Python optimization package for large scale quantum metrology with customized strategy structures}
\author{Piotr Dulian}
\affiliation{Centre for Quantum Optical Technologies, Centre of New Technologies, University of Warsaw, Banacha 2c, 02-097 Warszawa, Poland}
\affiliation{Faculty of Physics, University of Warsaw, Pasteura 5, 02-093 Warszawa, Poland}
\author{Stanis{\l}aw Kurdzia{\l}ek}
\affiliation{Faculty of Physics, University of Warsaw, Pasteura 5, 02-093 Warszawa, Poland}
\author{Rafa{\l}  Demkowicz-Dobrza{\'n}ski} 
\affiliation{Faculty of Physics, University of Warsaw, Pasteura 5, 02-093 Warszawa, Poland}

\begin{abstract}
QMetro++ is a Python package that provides a set of tools for identifying optimal estimation protocols that maximize quantum Fisher information (\acs{QFI}). Optimization can be performed for arbitrary configurations of input states, parameter-encoding channels, noise correlations, control operations, and measurements. The use of tensor networks and an iterative see-saw algorithm allows for an efficient optimization even in the regime of a large number of channel uses ($N\approx100$).
Additionally, the package includes implementations of the recently developed methods for computing fundamental upper bounds on \acs{QFI}, which serve as benchmarks for assessing the optimality of numerical optimization results. All functionalities are wrapped up in a user-friendly interface which enables the definition of strategies at various levels of detail.\\
\\
Link: \url{https://github.com/pdulian/qmetro}
\end{abstract}
\maketitle

\section{Introduction}

\begin{figure*}[t]
    \centering
    \includegraphics[width=0.9 \textwidth]{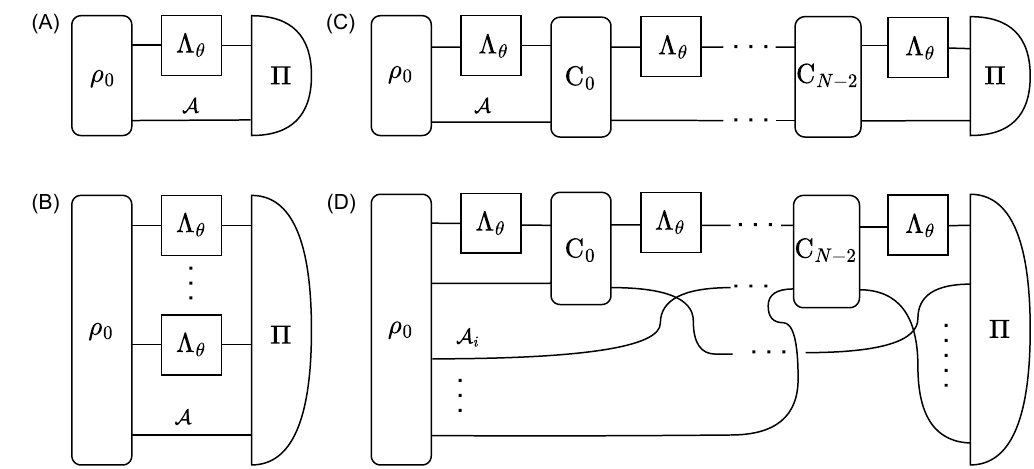}
    \caption{Different classes of metrological strategies that one can optimize using the package: (A) optimization of a \emph{single channel} input probe $\rho_0$ (potentially entangled with an ancillary system $\mathcal{A}$),  (B) optimization of an entangled state $\rho_0$ of $N$ input probes in a \emph{parallel} strategy (potentially additionally entangled with an ancillary system $\mathcal{A}$), (C) optimization of an input probe $\rho_0$ and control operations $C_i$ in an \emph{adaptive} metrological scheme, (D) a \emph{customized} protocol structure, here inspired by quantum collisional models, where multi-partite entangled ancillary state is sent piece-by-piece to interact with a common sensing system via interaction gates $C_i$. Here the optimization may affect either the input state $\rho_0$ or interaction gates $C_i$ or both. The above strategies correspond to functions listed in Table~\ref{tab:summary}.}  
    \label{fig:intro}
\end{figure*}
\begin{table*}[t!]
\centering
		\begin{tabular}{||c|c|c|c||}
			\hline 
			\multicolumn{2}{||c|}{\textbf{strategy}} &  \textbf{package functions} & \textbf{section} \\ \hline \hline
            \multicolumn{2}{||c|}{(A) single channel} & 
            \texttt{mop\_channel\_qfi}, \texttt{iss\_channel\_qfi} & \ref{sec:single}\\ \hline
             \multirow{3}{*}{\begin{tabular}{c}(B) \\ parallel \end{tabular}}  & small $N$ & \texttt{mop\_parallel\_qfi}, \texttt{iss\_parallel\_qfi} & \multirow{3}{*}{\ref{sec:parallel}}\\ 
               & large $N$ & \texttt{iss\_tnet\_parallel\_qfi} &\\ 
             & bound & \texttt{par\_bounds} &  \\ 
			\hline 
                & small $N$ & \texttt{mop\_adaptive\_qfi}, \texttt{iss\_adaptive\_qfi} & \multirow{4}{*}{\begin{tabular}{c}\ref{sec:adaptive}\\ \ref{sec:correlated}
                \end{tabular}}\\ 
             (C) & large $N$ & \texttt{iss\_tnet\_adaptive\_qfi} &  \\ 
             adaptive & bound & \texttt{ad\_bounds} &\\ 
             & corr. noise bound & \texttt{ad\_bounds\_correlated}& \\
			\hline 
             \multicolumn{2}{||c|}{(D) customized} & \texttt{iss\_opt} + \dots & \ref{sec:advanced}\\
             \hline
		\end{tabular}
		\caption{An overview of package functions that allow to identify the optimal metrological strategies with a particular protocol structure. Examples of  a particular protocol structure optimization, with complete Python codes, are provided in the sections listed in the last column. The corresponding structures are depicted in Fig.~\ref{fig:intro}.}
		\label{tab:summary}
\end{table*}

\begin{figure*}[p]
    \centering
    \includegraphics[width=1. \textwidth]{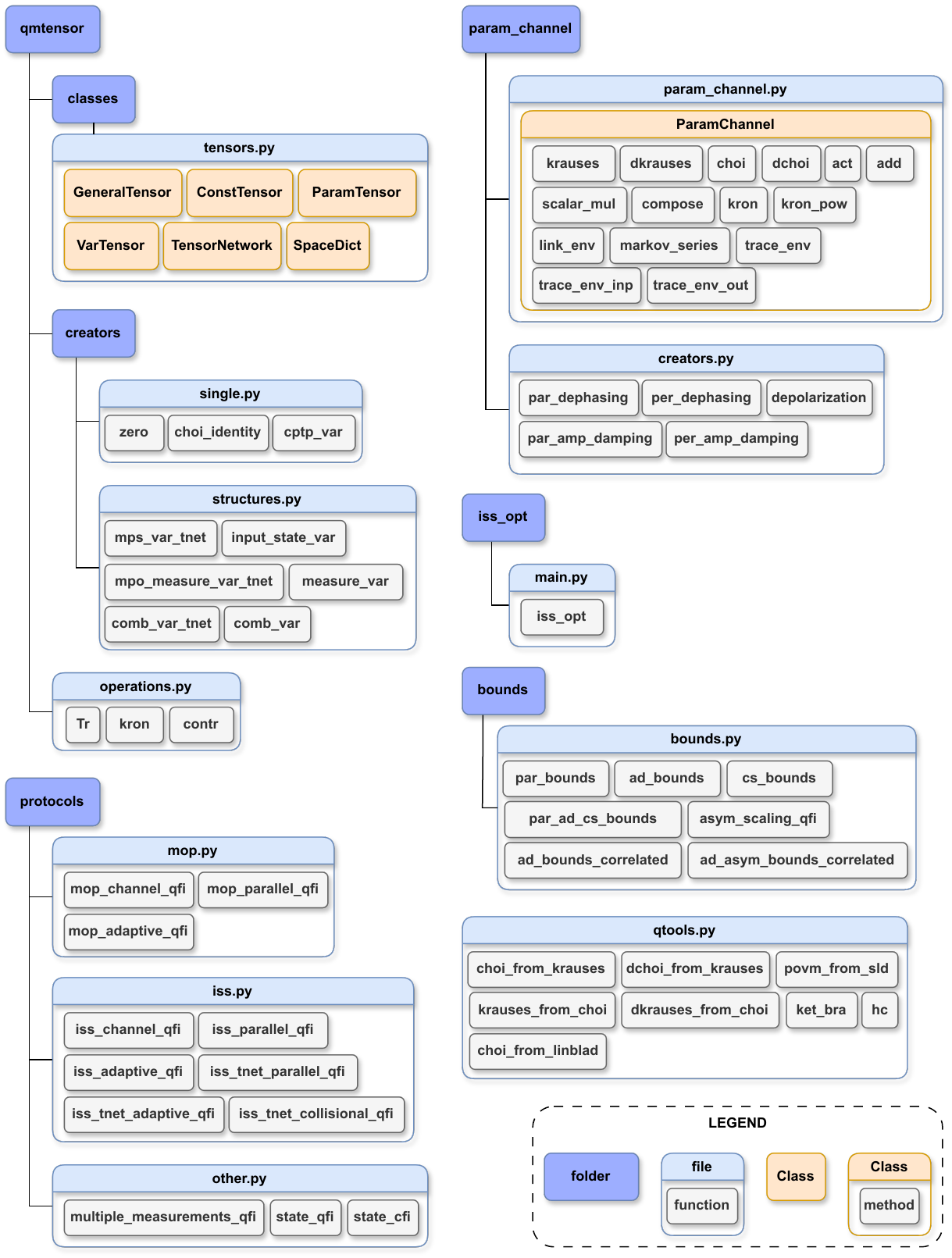}
    \caption{File structure of the package with a selection of the most important functions. Classes from \texttt{tensors.py} file are displayed in more detail in Fig.~\ref{fig:classes}. All functions, classes and methods are listed and described in the package documentation \cite{Dulian2025}.}
    \label{fig:scheme}
\end{figure*}

Quantum metrology \cite{Giovaennetti2006, Toth2014, PezzeSmerziOberthalerEtAl2018, Polino2020} is now rightfully regarded as one of the main pillars of the rapidly developing field of quantum technologies \cite{Simon2025}. It focuses on the ways to exploit delicate quantum features of light and matter to enhance the sensitivity of practical measuring devices. Apart from numerous experimental breakthroughs in the field, including the use of squeezed states of light in gravitational wave detectors \cite{ Schnabel2017, LIGO2023}, quantum-enhanced magnetometry \cite{Troullinou2023}, gravitometry \cite{Cassen2025}, atomic clocks \cite{PedrozoPeafiel2020} and many others, the field has also enjoyed significant theoretical progress with numerous analytical and numerical methods being developed that help
to design better metrological protocols as well as understand the fundamental limitations of quantum-enhanced protocols in the presence of decoherence \cite{Escher2011, Demkowicz2012, Demkowicz2014, demkowicz2015, Demkowicz2017, Zhou2018, Chabuda2020, Zhou2021, Altherr2021, Liu2023, Kurdzialek2023, kurdzialek2024, Liu2024}.

The purpose of this package is to make these advanced and powerful numerical methods accessible to the community. We realize that despite the fact that the core elements of the methods have been discussed in the literature already for a couple of years, they are still not widely used in the field due to their somewhat steep 'initial learning curve' and the fact that there is no single publication that provides a unified presentation of all of them, not to mention a numerical package. Moreover, we significantly expand the applicability of the latest tensor network-based methods to include general protocol structures, which allows for dealing with e.g. collisional metrological schemes.

Not so rarely, we have come across research conducted in the field, where the problem of optimization of quantum metrological protocols has been approached using general-purpose optimization tools (including even the latest AI methods), which in the end did not result in a particularly efficient numerical algorithm, the results of numerical optimization had no optimality guarantee 
and were not benchmarked against known fundamental precision bounds [citations intentionally hidden here!].

This package adopts the frequentist approach to quantum metrology and focuses on quantum Fisher information (\acs{QFI}) as the figure-of-merit to be optimized \cite{Helstrom1976, Braunstein1994, Paris2009}. 
At this point, we do not include quantum Bayesian methods in the package, even though some of the optimization procedures discussed here have their Bayesian counterparts, and may be added in the future updates. The main reason is that, from our numerical experience, the Bayesian approach 
does not combine well with the tensor-network optimization procedures, and hence cannot be efficiently used to study the performance of protocols in a large number of channel uses, see further discussion in the conclusion section.



The crucial and unique feature of our package is the ability to go to large-scale metrological problems, involving multiple channel uses, whether in adaptive sequential schemes, parallel schemes, or any conceivable protocol structure one may think of, and optimize the protocols without any prior assumption regarding the class of operations allowed, apart from the restriction on the dimensionality of the physical and ancillary spaces involved, see Fig.~\ref{fig:intro}.

As such, our package completely supersedes the TNQMetro package \cite{Chabuda2022},
which was designed to deal with optimization of input probes in a parallel metrological scheme involving a large number of channels. Compared with the previous package, the new one not only allows for a full flexibility   
of choosing the protocol structure one wants to analyze, but even in the basic parallel channels scenario offers a significantly improved user-friendly interface that allows to deal with a large class of problems, including correlated noise models, in a straightforward way.

The second unique component of the package is the implementation of 
the methods that compute fundamental metrological bounds in a very efficient way. These bounds are the tightest universal bounds known in the literature \cite{Kurdzialek2023} and include the very latest method to compute bounds in case of \emph{correlated} noise models as well \cite{kurdzialek2024bounds}! The bounds may serve as benchmarks for the protocols found as a result of numerical optimization. 

Interestingly, in case of uncorrelated noise models, it is known that the bounds are tight in the asymptotic limit of large number of channel uses \cite{Zhou2021, Kurdzialek2023}. This fact makes the bounds powerful in two ways. Not only do they serve as a proof that a given protocol \emph{is optimal}, if it saturates the bound, but also indicate that the protocol is in fact \emph{not optimal} if one sees a persistent gap between the performance of the numerically found protocol and the bound that does not diminish with the increase of the number of channels used. 

The main functionalities of the package are summarized in Table~\ref{tab:summary}, where we indicate the methods  designed to optimize the strategies corresponding to the structures depicted in Fig.~\ref{fig:intro}, including single-channel strategy (A), parallel channels strategy (B), adaptive strategy (C) and customized strategy (D).

In the small-$N$ regime,  protocols (A), (B) and (C) may be optimized as a whole, without the use of tensor-network structures. This allows us to perform the optimization using two alternative approaches: the iterative see-saw (\acs{ISS})  and the minimization over purifications (\acs{MOP}).  

The \acs{MOP} approach has the advantage over \acs{ISS} that the numerical solution found is guaranteed to be optimal, see Sec.~\ref{sec:mop}. Hence, this is the method to be chosen for small-scale problems, where dimensions of the Hilbert spaces are reasonably small. However, \acs{MOP} methods cannot be generalized to large-scale problems and combined with tensor-network framework---then, one needs to resort to the \acs{ISS} approach. The \acs{ISS} has no guarantee to converge to the true optimal solution, but is constructed in a way that every iteration step cannot reduce the optimized figure-of-merit, see Sec.~\ref{sec:iss}.

The main advantage of \acs{ISS} over \acs{MOP} is that it naturally combines with tensor-network formalism and may work in the large-$N$ regime. In this case the optimization is performed over each local node in the tensor-network structure separately, whether it represents a part of the input state, control operation or measurement. For this approach one may use the higher-level functions, \texttt{iss\_tnet\_parallel\_qfi} or \texttt{iss\_tnet\_adaptive\_qfi}, dedicated to optimizing standard protocol structures such as parallel (B) or adaptive (C) ones respectively. Alternatively, one can construct a customized structure of the protocol, and use a lower-level function \texttt{iss\_opt} which performs the iss optimization for arbitrary tensor-network structure, see Sec.~\ref{sec:advanced} for a detailed example.  

Finally, the package comes with a number of functions that allows one to compute fundamental bounds given just the structure of an individual parameter-encoding quantum channel. 
The two functions, \texttt{par\_bounds} and \texttt{ad\_bounds} allow to compute the bounds in case of uncorrelated noise models valid for parallel and adaptive strategies. These bounds are known to be asymptotically saturable for $N\rightarrow \infty$, see Sec.~\ref{sec:bounds}. The function \texttt{ad\_bounds\_correlated}, on the other hand, provides bounds in case of correlated noise models, where there is some quantum environment that connects different instances of channels action, and captures the effects of e.g. non-Markovianity. This function provides valid bounds which may be tightened at the expense of numerical complexity, but unlike uncorrelated noise bounds, they are not guaranteed to be tight.

The detailed structure of the package is depicted in Fig.~\ref{fig:scheme}. The higher-level optimization functions are collected in the \texttt{protocols} folder, including both \acs{MOP} and \acs{ISS} protocols, with the latter one in the standard as well as the tensor-network variant (\texttt{tnet}). This folder also provides a couple of auxiliary functions: a function for computation of \acs{QFI} of a state (\texttt{state\_qfi}), a function for computation of classical Fisher information given a state and measurement (\texttt{state\_cfi}) and a function \texttt{multiple\_measurements\_qfi} that calculates the optimal distribution of $N$ channels into $k$ experiments with $N_i$ ($\,N= \sum^k_{i=1} N_i$) channels.
The functions computing bounds are gathered in the \texttt{bounds} folder, where apart from the standard procedure to compute the bounds for finite number of channel uses, there are also asymptotic variants that compute the bounds in the asymptotic limit $N \rightarrow \infty$. 

As a bonus feature, we have also included bounds that are valid even in the more general than adaptive \emph{causal superpositions} strategies (\texttt{cs}), see \cite{Kurdzialek2023}. They are, however, not the essential element of the package, since they do not have a counterpart large-scale optimization procedure to find the optimal protocols, as tensor-network formalism is not obviously compatible  with this framework.

In order to perform the optimization using higher-level functions or compute the bounds, one needs to specify the form of the parameter-dependent channel $\Lambda_\theta$. The \texttt{param\_channel} folder contains all the necessary methods to construct the channel and includes also some 
exemplary predefined channels in the \texttt{creators.py} file.
The channel may be provided in a number of different ways: either through a list of Kraus operators, or a \acs{CJ}  matrix, or finally as a time integrated quantum Master equation, see Sec.~\ref{sec:single}.

For optimization of protocols with customized structures one needs to resort to low-level \acs{ISS} optimization function \texttt{iss\_opt} and  construct the desired protocol structure using the classes provided in \texttt{tensors.py} file, which allow to specify the elements of the protocol that will be optimized (\texttt{VarTensor}), elements that are fixed (\texttt{ConstTensor}) and elements that include parameter to be estimated (\texttt{{ParamTensor}}), see Sec.~\ref{sec:advanced}.

The package comes also with a number of useful auxiliary functions, combined in \texttt{qtools.py} module. Among them functions that allow to switch between different representations of quantum channels like Kraus, Choi-Jamio{\l}kowski and Lindbladian operators. We describe them throughout Sec.~\ref{sec:basic} and \ref{sec:advanced}.

This paper is organized as follows. In Sec.~\ref{sec:theory} we provide a theoretical background for the essential quantum metrological concepts the package relies on. We formulate the quantum metrological problem as a quantum channel estimation, and introduce the \acs{QFI} as the figure of merit to be optimized. Further on, we introduce both \acs{MOP} and \acs{ISS} optimization procedures. Finally, we briefly review the methods to compute fundamental bounds.

Sec.~\ref{sec:basic} aims to guide the reader through the 
high-level optimization functions, going step by step: explaining first how to provide the form of the parameter-dependent channel, how to perform the simplest single-channel optimization problems, how to deal with the optimization of parallel and adaptive strategies and finally how to optimize metrological protocols in the most challenging scenarios where noise correlations between subsequent channel uses are present. 

Sec.~\ref{sec:advanced} complements the previous one, and shows the full potential of the package using the example of \emph{collisional} metrological scenario, where one needs to design a customized protocol tensor-network structure from the basic building blocks, and use the low-level \texttt{iss\_opt} optimization function. 

We summarize the presentation  in Sec.~\ref{sec:summary} indicating some limitations and further possible developments of the package.

\section{Theoretical background}
\label{sec:theory}
\subsection{Quantum channel estimation problem}\label{sec:qmetro}
In a paradigmatic quantum metrological problem, the goal is to estimate the value of a single parameter $\theta$ encoded in some physical process described by a quantum channel (formally a Completely Positive Trace-Preserving map -- \acs{CPTP}):
\begin{equation}
    \Lambda_\theta: \L{\mc I} \mapsto \L{\mc O},
\end{equation}
where $\mc I$, $\mc O$ are Hilbert spaces of input and output quantum systems and $\L{\mathcal{H}}$ is a set of linear operators on $\mathcal{H}$, that is a superset of density matrices on $\mathcal{H}$. 

The channel $\Lambda_\theta$ can be probed by an arbitrary $\theta$-independent input state \begin{equation}
    \rho_0 \in \L{\mc I \otimes \mathcal{A}},
\end{equation}
where  $\mathcal{A}$ is an auxiliary system called an \emph{ancilla};  $\Lambda_\theta$ does not act on $\mathcal{A}$, but the entanglement between $\mathcal{I}$ and $\mathcal{A}$ may sometimes be used to enhance the estimation precision in the presence of noise \cite{Fujiwara2001, Kolodynski2013, kurdzialek2024, Liu2024}. The resulting output state is
\begin{equation}
    \rho_\theta := \left( \Lambda_\theta \otimes \mathbb{I}_{\mathcal{A}} \right)(\rho_0) \in \L{\mc O \otimes \mathcal{A}},
\end{equation}
 where $\mathbb{I}_\mathcal{A}$ represents the identity map on $\mathcal{L}(\mathcal{A})$---note the notational distinction from $\bbone_{\mathcal{A}}$, which denotes identity operator on the $\mathcal{A}$ space. 
 
 The state $\rho_\theta$ is measured by a generalized measurement $\Pi=\{\Pi_i\}$ resulting in the probability of outcome $i$ equal to $p_\theta(i) := \Tr\left(\rho_\theta \Pi_i\right)$.
The parameter $\theta$ is then estimated using an estimator  $\tilde\theta(i)$ and the goal is to find the protocol producing results as close as possible to the true value of $\theta$. In case of the simplest single-channel estimation strategy, as depicted in Fig.~\ref{fig:intro}(A), the optimization of the protocol is equivalent to optimizing the input state $\rho_0$ and the measurement $\{\Pi_i\}$.
However, more complex protocols may additionally involve optimization over control operations $\t{C}_i$, as in scenarios (C) and (D) in Fig.~\ref{fig:intro}. 

In general, the exact form of the optimal protocol depends on the actual estimation framework chosen, be it frequentist or Bayesian \cite{Hayashi2011, Meyer2023, DemkowiczDobrzanski2020}, as well as the choice of the cost function/figure of merit to be minimized/maximized.  In this work, we adopt the frequentist approach, and focus on optimizing the \acs{QFI} as the figure of merit to be maximized.

\subsection{Quantum Fisher Information}
\label{sec:fisher}

Given some probabilistic model that specifies probabilities $p_\theta(i)$ of obtaining different measurement outcomes $i$ depending on the value of a parameter $\theta$, one can lower-bound
the achievable mean squared error $\Delta^2 \tilde{\theta}$ of any locally unbiased estimator $\tilde{\theta}(i)$ 
via the famous \emph{classical Cram{\'e}r-Rao}  (\acs{CCR}) bound \cite{kay1993}:
\begin{equation}\label{eq:cr_bound}
    \Delta^2\tilde\theta \ge \frac{1}{k F_C(p_{\theta})}, 
\end{equation}
where $k$ is the number of independent repetitions of an experiment and $F_C$ is \emph{classical Fisher information} (\acs{CFI}):
\begin{equation}\label{eq:cfi}
    F_C(p_{\theta}) := \sum_i \frac{\dot p_\theta(i)^2}{p_\theta(i)}, 
\end{equation}
where $\dot{p}_\theta$ denotes the derivative with respect to $\theta$.
The bound is asymptotically saturable, in the limit of 
many repetitions of an experiment $k \rightarrow \infty$, with the help of e.g. a maximum-likelihood estimator \cite{kay1993}.
Hence, the larger the \acs{CFI}, the better parameter sensitivity of the model. 

In quantum estimation models, the outcomes are the results of the measurement $\{\Pi_i\}$ performed on a quantum state $\rho_\theta$,  $p_\theta(i) = \Tr\left(\rho_\theta \Pi_i \right)$. Hence, depending on the choice of the measurement, one may obtain larger or smaller \acs{CFI}. 
The \acs{QFI} corresponds to the maximal achievable \acs{CFI}, when  optimized over all admissible quantum measurements \cite{Helstrom1976, Braunstein1994, Paris2009}:
\begin{equation}\label{eq:qfi}
    F_Q(\rho_{\theta}) := \max_{\{ \Pi_i \}_i} F_C(p_{\theta}) =  \Tr\left(\rho_{\theta} L^2\right),
\end{equation}
where $L$ is the \emph{symmetric logarithmic derivative} (\acs{SLD}) of $\rho_\theta$:
\begin{equation}
    \dot \rho_\theta = \frac{1}{2}\left( \rho_{\theta}L + L\rho_{\theta}
    \right).
\end{equation}
Moreover, the optimal measurement can, in particular, be chosen as the projective measurement in the eigenbasis of the \acs{SLD}.

Combining the above with the \acs{CCR} bound \eqref{eq:cr_bound} leads to the \emph{quantum Cram{\'e}r-Rao} (\acs{QCR}) bound that sets a lower bound on achievable estimation variance when estimating a parameter encoded in a quantum state $\rho_\theta$:
\begin{equation}
    \Delta^2 \tilde{\theta} \geq \frac{1}{k F_Q(\rho_\theta)},
\end{equation}
and is saturable in the limit of $k \rightarrow \infty$.

This makes the \acs{QFI} a fundamental concept in quantum estimation theory, and makes it meaningful to formulate optimization problems in quantum metrology in the form of an optimization of the \acs{QFI} of the final quantum state obtained at the output of a given metrological protocol.

The basic single-channel estimation task, depicted in Fig.~\ref{fig:intro}(A), corresponds then to the following optimization problem 
\begin{equation}
\label{eq:channelqfi}
F_Q(\Lambda_\theta ) = \max_{\rho_0} F_Q \left[\Lambda_\theta\otimes \mathbb{I}_\mathcal{A}(\rho_0)\right].
\end{equation}
We will refer to $F_Q(\Lambda_\theta)$ as the \emph{channel} \acs{QFI}, as it corresponds to the maximal achievable \acs{QFI} of the output state of the channel, when the maximization over all input states is performed. 

In case of protocols with multiple channel uses, $N > 1$, one could simply probe each channel independently, using independent probes and independent measurements. In this case, by additivity of \acs{QFI}, one would end up with the corresponding $N$-channels \acs{QFI} --- $F_Q^{(N)}(\Lambda_\theta)$ equal, for this simplistic strategy, to $N$ times the single channel \acs{QFI}.

However, one may also consider more sophisticated strategies involving multiple channel uses as depicted in Fig.~\ref{fig:intro}(B),(C),(D), 
which may provide significant advantage over the independent probing strategy, in principle leading even to quadratic scaling of \acs{QFI} --- $F_Q^{(N)}(\Lambda_\theta)\propto N^2$, referred to as the \emph{Heisenberg scaling} (\acs{HS}) \cite{Giovaennetti2006}. 

In case of noiseless unitary estimation models $
\Lambda_\theta(\rho) = U_\theta \rho U_\theta^\dagger$ the optimal probe states that maximize the \acs{QFI} are 
easy to identify analytically \cite{Giovaennetti2006} and no sophisticated tools are required. In case of noisy models, however, the actual optimization over the protocol is much more involved, as in principle it requires optimization over multipartite entangled states (B) or multiple control operations  $\textrm{C}_i$ (C)
or both (D). 

This package allows for optimization of all kinds of metrological strategies, employing ancillary systems, entanglement and control operations, with the only physical restriction that the operations applied have a definite causal order, and hence can be regarded as so-called \emph{quantum combs} in the sense formalized in \cite{Chiribella2009} and further in this text. 

Regarding optimization strategies, we start with the discussion of \acs{MOP} method, which is applicable to small-scale scenarios, and then move on to discuss a more versatile \acs{ISS} optimization which can be combined with tensor-network formalism to deal with large-scale metrological problems. 

\subsection{Minimization over purifications}\label{sec:mop}
The \acs{MOP} method \cite{Fujiwara2008, Escher2011, Demkowicz2012, kurdzialek2024} is based on an observation that the \acs{QFI} of a state $\rho_\theta\in\mc L(\mc H)$ is equal to the minimum of \acs{QFI} over its purifications:
\begin{equation}
    F_Q(\rho_\theta) = \min_{\ket{\Psi_\theta}}F_Q(\ket{\Psi_\theta}\bra{\Psi_\theta}),
\end{equation}
where $\ket{\Psi_\theta} \in \mc H \otimes \mc R$ is a purification of $\rho_\theta, \, \rho_\theta=\Tr_\mc{R}\ket{\Psi_\theta}\bra{\Psi_\theta}$. 

Consider now a parameter-encoding channel  $\Lambda_\theta: \mathcal{L}(\mathcal{I}) \mapsto \mathcal{L}(\mathcal{O})$. 
The idea of minimization over purifications leads to an efficiently computable formula for the channel \acs{QFI} \cite{Fujiwara2008, Escher2011, Demkowicz2012, kurdzialek2024} 
\begin{equation}
\label{eq:minkraus}
    F_Q(\Lambda_\theta) = 4\min_{\{K_{\theta, k}\}} \left\|
    \sum_{k=1}^r \dot{K}_{\theta,k}^\dagger \dot{K}_{\theta,k}
    \right\|,
\end{equation}
where $\| \cdot \|$ is the operator norm and the minimization is performed over all equivalent Kraus representations of the channel, such that:
\begin{equation}
    \Lambda_\theta(\cdot)= \sum_{k=1}^r K_{\theta,k} \cdot K_{\theta,k}^\dagger.
\end{equation}
This optimization becomes feasible, as one can effectively parametrize the derivatives of the Kraus operators corresponding to all equivalent Kraus representations as:
\begin{equation}
\dot K_{\theta, k}(h) = \dot{K}_{\theta, k}-\I \sum_l h_{kl}  K_{\theta, l},
\end{equation}
where $h \in \mathbb{C}^{ r\times r}$ is a Hermitian matrix and $K_{\theta,k}$ is any fixed Kraus representation of $\Lambda_\theta$, e.g. the canonical representation obtained from eigenvectors of the corresponding Choi-Jamio{\l}kowski (\acs{CJ}) operator \cite{Bengtsson2006}, see Appendix \ref{sec:link_prod} for the definition of the \acs{CJ} operator. 

As a result, one may rephrase the apparently difficult minimization problem \eqref{eq:minkraus} as:
\begin{equation}
    F_Q(\Lambda_\theta) = 4 \min_{h} \| \alpha(h)\|, 
\end{equation}
where 
\begin{equation}
\label{eq:alpha}
 \alpha(h) = \sum_{k}
    \dot K_{\theta, k}(h)^\dagger \dot K_{\theta, k}(h).
\end{equation}
Importantly, this problem may be effectively written as a simple semidefinite program (\acs{SDP}) \cite{Demkowicz2012, kurdzialek2024} 
\begin{equation}
\label{eq:mop_sdp}
    F_Q(\Lambda_\theta) = 4 \min_{\substack{\lambda \in \mb R,\\ h=h^\dagger}} \lambda  ~~\textrm{s.t.}~~ A \succeq 0,
\end{equation}
where
\begin{equation}
A = \left( \begin{array}{c|ccc}
        \lambda \mathbb{1}_\mathcal{I}& \dot{{ K}}_{\theta,1}^\dagger(h)  & \hdots &  \dot{{ K}}_{\theta,r}^\dagger(h) \\ \hline
        \dot{{ K}}_{\theta,1}(h)&  &  &  	 \\
        \vdots&  & \mathbb{1}_{d \cdot r} &    \\ 
        \dot{{ K}}_{\theta, r}(h) &  &  &            
 \end{array}\right), 
\end{equation}
$d= \textrm{dim} \mathcal{O}$ and $r$ is the number of Kraus operators.
Provided the dimension of the relevant Hilbert space is small enough, this problem may be solved efficiently using widely available \acs{SDP} solvers, such as \cite{mosek}. This optimization is implemented in the  \texttt{mop\_channel\_qfi} function. 

\begin{figure*}[t]
\centering
    \includegraphics[width=0.9 \textwidth]{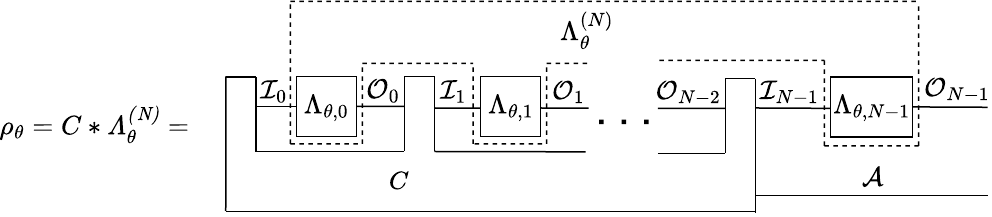}.
\caption{Final state resulting from the action of an adaptive protocol on $N$ parameter encoding channels, written via a formal link product operation between the corresponding \acs{CJ} operators---quantum combs.}
\label{fig:linkadaptive}
\end{figure*}

Notice that the dimension of ancilla $\mc A$ is not included in any of the constraints in \eqref{eq:mop_sdp}. This is because \acs{MOP} gives the maximal \acs{QFI}  without any restriction on  ancilla sizes.
Furthermore, in order to identify the optimal input state $\ket\psi$ one needs to solve an additional \acs{SDP} problem, see \cite{Zhou2021, kurdzialek2024}.

\acs{MOP} can also be used to optimize \acs{QFI} in case of multiple channel uses. For the parallel strategy, Fig.~\ref{fig:intro}(B), one can simply apply the single-channel method, replacing the channel $\Lambda_\theta$ with $\Lambda_\theta^{(N)}=\Lambda_\theta^{\otimes N}$---this is implemented in the $\texttt{mop\_parallel\_qfi}$ function. 


In order to address an adaptive strategy, as in Fig.~\ref{fig:intro}(C), one should first use the fact that a concatenated sequence of quantum channels  $\mr C_i$ where the free inputs and free output spaces are 
respectively $\mathcal{H}_{2i}$, $\mathcal{H}_{2i+1}$, $i \in {0, \dots, N-1}$, can be represented as a quantum comb \cite{Chiribella2009}
\begin{equation}
C\in \mr{Comb}[(\mc H_0, \mc H_1), ..., (\mc H_{2N-2}, \mc H_{2N-1})],
\end{equation}
which is a \acs{CJ} operator satisfying:
\begin{align}\label{eq:comb_cond}
    &C \succeq 0, \quad C^{(N)}=C, \quad C^{(0)}=1, \\ \nonumber
    &\Tr_{2k-1}C^{(k)} = \bbone_{2k-2} \otimes C^{(k-1)}, \quad k\in \{1, ..., N\}.
\end{align}
The above conditions guarantee that such quantum comb $C$ can always be regarded as a concatenation of quantum channels $C_i$, namely $C=C_0*...*C_{N-1}$, where $*$ denotes the link product \cite{Chiribella2009}, see Appendix~\ref{sec:link_prod} for the definition of the link product.

In the context of metrological adaptive scenario, 
given a series of channels $\Lambda_{\theta, i}: \mc L(\mc I_i)\mapsto \mc L(\mc O_i)$, we can write the combined action of all the channels as:
\begin{equation}
\label{eq:uncorchan}
\Lambda^{(N)}_\theta=\Lambda_{\theta, 0}\otimes ... \otimes \Lambda_{\theta, N-1}.
\end{equation}
We may now represent the general adaptive strategy that includes the input state as well as the control operations in the form of a quantum comb
\begin{multline}
C \in \mr{Comb}[(\mc \emptyset, \mc I_0), (\mc O_0, \mc I_1), ..., (\mc O_{N-3}, \mc I_{N-2}),\\ (\mc O_{N-2}, \mc I_{N-1}\otimes\mc A)]
\end{multline}
and concatenate the operations using the link product in order to obtain  the output state 
\begin{equation}
    \rho_\theta = C * \mathit{\Lambda}_{\theta}^{(N)} \in \mathcal{L}(\mathcal{O}_{N-1}\otimes \mathcal{A}),
\end{equation}
where $\mathit{\Lambda}_{\theta}^{(N)}$ (italics) formally represents the \acs{CJ} operator of $\Lambda_{\theta}^{(N)}$, see Fig.~\ref{fig:linkadaptive}. Here $\mathit{\Lambda}^{(N)}_\theta$
does not necessarily need to represent the action of uncorrelated channels as in \eqref{eq:uncorchan}, but may also be a general parameter-dependent quantum comb and in this way represent models with correlated noise or non-local character of estimated parameter.  

Optimization over adaptive protocols is thus equivalent to maximizing the final \acs{QFI} over quantum combs $C$. 
Using again the \acs{MOP} idea, this problem may be effectively written down as an \acs{SDP} (this time slightly more involved) of the following form 
\cite{Altherr2021, Liu2023, kurdzialek2024}:
\begin{align}
&F_{\t{Q}}^{(N)}\left(\Lambda_\theta\right) = 4 \min_{\substack{\lambda \in \mb R, Q^{(k)}\\ h=h^\dagger}} \lambda  ~~\textrm{s.t.}~~ A \succeq 0,  \\
    \nonumber
&\underset{2\leq k \leq N-1}{\forall}\t{Tr}_{\mathcal{O}_{k-1}}Q^{(k)} = \openone_{\mathcal{I}_{k-1}} \otimes Q^{(k-1)}, \; Q^{(0)}= 1,    
    \end{align}
where  $Q^{(k)} \in \mathcal{L}(\mathcal{O}_0 \otimes \mathcal{I}_0 \otimes \dots \otimes \mathcal{O}_{k-1} \otimes \mathcal{I}_{k-1})$ are optimization variables and $A=$
\begin{equation}
 \left( \begin{array}{c|c}
        \openone_{\mathcal{I}_{N-1}} \otimes Q^{(N-1)}& \ket{\dot{\tilde{K}}_{1,1}(h)}  \hdots  \ket{\dot{\tilde{K}}_{r,d}(h)} \\ \hline
        \bra{\dot{\tilde{K}}_{1,1}(h)}&    	 \\
        \vdots&   \lambda \mathbb{1}_{d  r}      \\ 
       \bra{\dot{\tilde{K}}_{r,d}(h)} &             
 \end{array}\right)
\end{equation}
with 
\begin{equation}
    \ket{\dot{\tilde{K}}_{i,k}(h)} = {_{\mathcal{O}_{N-1}}}\!\!\braket{i|\dot{K}_{\theta,k}^{(N)}(h)}, 
\end{equation}
where 
\begin{equation}
    \ket{\dot{{K}}_{i,k}^{(N)}(h)}  \in \mathcal{O}_0 \otimes \mathcal{I}_0 \otimes \dots \otimes \mathcal{O}_{N-1} \otimes \mathcal{I}_{N-1},
\end{equation}
are derivatives of vectorized Kraus operators of $\Lambda_{\theta}^{(N)}$, $r$ is the number of Kraus operators of $\Lambda_\theta^{(N)}$, $d = \textrm{dim} \mathcal{O}_{N-1}$.
This optimization is implemented in \texttt{mop\_adaptive\_qfi} function.

Similarly to the single channel case, optimization of parallel and adaptive strategies with \acs{MOP} does not allow for the control of the ancilla dimension and finding optimal input state or optimal comb requires solving an additional \acs{SDP} problem \cite{Liu2023, kurdzialek2024}.
Moreover, since the algorithm requires optimization over Kraus representations of $\Lambda^{(N)}_\theta$ its time and memory complexity is exponentially large in $N$.

\subsection{Iterative see-saw}\label{sec:iss}

In the \acs{ISS} optimization method developed in the context of quantum metrology \cite{Demkowicz2011, Macieszczak2013, Macieszczak2014, Toth2018, Chabuda2020, Geza2022, kurdzialek2024} one considers the \emph{pre-QFI function}:
\begin{equation}\label{eq:preqfi}
    F(\rho_0, \mf L) := 2\Tr \left( \dot \rho_\theta \mf L \right) - \Tr \left(\rho_\theta \mf L^2 \right),
\end{equation}
where $\rho_\theta = \Lambda_\theta \otimes \mathbb{I}_\mathcal{A} (\rho_0)$, and the dimension of $\mathcal{A}$ can be chosen arbitrarily and this choice may affect the final result, unlike in the \acs{MOP} method. It turns out that for a given $\rho_0$ the maximization of $F$ over Hermitian matrices $\mf L$ yields the \acs{QFI} of the state $\rho_\theta$:
\begin{equation}
    \max_{\mf L=\mf L^\dagger}F(\rho_0, \mf L) = F_Q(\rho_\theta),
\end{equation}
with the solution $\mf L^\diamond$ equal to the \acs{SLD} matrix $L$ of $\rho_\theta$ \cite{Macieszczak2013}---hence we will refer to $\mf L$ as the \emph{pre-SLD matrix} for $\mf L$. It follows that the channel \acs{QFI} \eqref{eq:channelqfi} is simply a maximization of $F$ over both of its arguments:
\begin{equation}
    F_Q(\Lambda_\theta) = \max_{\rho_0, \mf L}F(\rho_0, \mf L).
\end{equation}
This problem can be solved by initializing $\rho_0$ and $\mf L$ at random and then by alternately making one argument constant and maximizing the other one. Each such optimization will increase the value of $F$ and we can proceed until the change of $F$ in a number of consecutive iterations becomes smaller than some established precision $\epsilon$. In case of the single-channel optimization the procedure is implemented in \texttt{iss\_channel\_qfi}.

\acs{ISS} algorithm can be easily applied to the parallel strategy by simply substituting $\Lambda_\theta$ with $\Lambda^{(N)}_\theta = \Lambda^{\otimes N}_\theta$---this is implemented in \texttt{iss\_parallel\_qfi} function. This approach, however, suffers from the same problem as \acs{MOP}---namely, it has an exponential  complexity in $N$. 

The solution is to split $\rho_0$ and $\mf L$ into smaller parts by expressing them as tensor networks (see Appendix \ref{sec:tensor_networks}). Let us set $\mc I^{(N)} = \bigotimes_{i=0}^N\mc I_i$ and $\mc O^{(N)} = \bigotimes_{i=0}^N\mc O_i$, where for $i< N$ spaces $\mc I_i$ and $\mc O_i$ are inputs and outputs of the $i$-th channel $\Lambda_{\theta, i}$ and $\mc I_N=\mc O_N=\mc A$. Then a state
\begin{equation}
    \ket{\psi} = \sum_{a_0,..., a_N}\psi^{a_0...a_N}\ket{a_0...a_N} \in \mc I^{(N)},
\end{equation}
can be expressed as a tensor network called a \emph{matrix product state} (\acs{MPS}) \cite{Bridgeman2017, Schollwock2011}:
\begin{equation}\label{eq:mps}
    \includegraphics[scale=0.75]{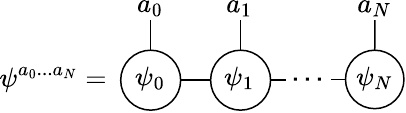},
\end{equation}
where indices $a_i$ are called \emph{physical indices} and horizontal indices that connect tensors $\psi_i$ are called \emph{bond indices}.
Range of the bond indices, called the \emph{bond dimension} and denoted by $r_\mr{MPS}$, controls the strength of possible correlations between sites of $\ket{\psi}$,  $r_\mr{MPS}=1$ means that the state is separable and $r_\mr{MPS}\approx\dim\mc I^{(N)}$ allows for arbitrary correlations.
Analogously, pre-SLD matrix can be expressed as a \emph{matrix product operator} (\acs{MPO}):
\begin{equation}\label{eq:sld_mpo}
    \includegraphics[scale=0.75]{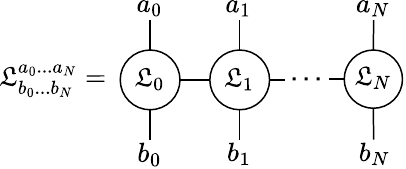}.
\end{equation}
with bond dimension $r_\mf{L}$.
We can then write the pre-QFI function as
\begin{gather}
    F(\psi_0, ..., \psi_N, \mf L_0, ..., \mf L_N) = \Asterisk_{i=0}^{N} \rho_0^i *\\\nonumber
     *\left[2\dot{ \mathit{\Lambda}}_{\theta}^{(N)} * \left( \Asterisk_{i=0}^{N} \mf L_i \right)^T  - \mathit{\Lambda}_{\theta}^{(N)} * \left( \Asterisk_{i=0}^{N} \mf L_i^2 \right)^T\right],
\end{gather}
where $\rho_0^i$ are elements of the density matrix \acs{MPO} of the state $\ket{\psi}$:
\begin{equation}\label{eq:mps_density_mat}
    \includegraphics[scale=0.75]{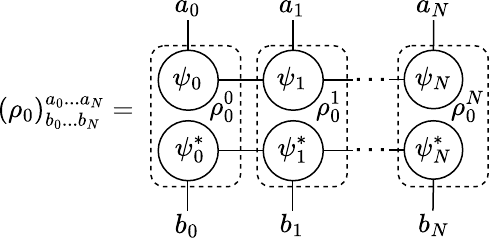},
\end{equation}
$\psi_i^*$ are entry-wise complex conjugations of $\psi_i$ and $\mf L_i^2$ are elements of \acs{MPO} of $\mf L^2$. Additionally, we used linearity of link product and the identity $\Tr(AB) = A*B^T$. Note that $\ast$ denotes link product operation, but also contraction of corresponding spaces for more general tensor networks, e.g. \acs{MPS}s, see Appendix~\ref{sec:tensor_networks} for more details. 

Just like in the case of a single channel optimization, we can maximize the above function over one argument at a time while others stay constant. Then after optimization of each argument, we repeat the procedure until the value of $F$ converges. Note that if $r_\mf{L}$ is too small, then the optimization over pre-SLD will not converge to the \acs{SLD} matrix and value of pre-QFI will not be equal to \acs{QFI}. Nevertheless, the obtained value of $F$ can always be interpreted as a \acs{CFI} for a measurement given by eigenvectors of $\mf L$ \cite{Macieszczak2013}.
The procedure employing \acs{MPS} and \acs{MPO} to the optimization of parallel strategy is implemented in \texttt{iss\_tnet\_parallel\_qfi}.

For the adaptive strategy we can apply the formula on $\rho_\theta$ from Fig.~\ref{fig:linkadaptive} to \eqref{eq:preqfi} and redefine pre-QFI function as
\begin{equation}\label{eq:preqfi_ad}
    F(C, \mf L) = 2 C * \dot{ \mathit{\Lambda}}_{\theta}^{(N)} * \mf L^T  - C * \mathit{\Lambda}_{\theta}^{(N)} * (\mf L^2)^T.
\end{equation}
Then, the \acs{QFI} for adaptive strategy becomes a maximization of \eqref{eq:preqfi_ad} over Hermitian $\mf L$ and quantum combs $C$ \eqref{eq:comb_cond}---this is implemented in \texttt{iss\_adaptive\_qfi}. Once more, this problem can be solved more efficiently if we split arguments of $F$ into smaller parts. This time, however, the tensor network of $\t{C}$ has a natural interpretation as an input state $\rho_0$ and a series of control operations $\mr C_i$ like in Fig.~\ref{fig:intro}(C). Thus we can optimize
\begin{multline}\label{eq:preqfi_ad_tnet}
    F(\rho_0, C_0, ..., C_{N-2}, \mf L) =\\\nonumber
    = \rho_0 * \Asterisk_{i=0}^{N-2} C_i * \left[2\dot{ \mathit{\Lambda}}_{\theta}^{(N)} * \mf L^T  - \mathit{\Lambda}_{\theta}^{(N)} * (\mf L^2)^T\right].
\end{multline}
We implemented this procedure in \texttt{iss\_tnet\_adaptive\_qfi}.

Finally, \acs{ISS} can be applied to any arrangement of input states, control operations and parametrized channels, see the example in Fig.~\ref{fig:intro}(D). This is done by expressing $\rho_\theta$ as the contraction of tensors representing all channels, measurements (possibly represented by \acs{MPO}s) and states (possibly represented by \acs{MPS}s) and maximizing pre-QFI over each variable node separately.  The package provides tools to define such arbitrary protocols which can be then optimized using \texttt{iss\_opt} function. We discuss this in detail in Sec.~\ref{sec:advanced}.

Let us conclude this section with a comparison of \acs{MOP} and \acs{ISS}. \acs{MOP} is a deterministic algorithm with optimality of the numerical solution guaranteed by formal proofs. \acs{ISS} initializes the parameters of pre-QFI function at random, thus it is nondeterministic. Additionally, there is no guarantee that it converges to the true optimum of the pre-QFI function. The optimality can be validated in some cases by comparing its result with \acs{MOP} or upper bounds, see Sec.~\ref{sec:bounds}. It might be also useful to repeat the computation several times and check whether the returned values vary. 

On a positive side, what is guaranteed in \acs{ISS} is that the figure of merit will not decrease at any individual iteration step and that the final result will always be a correct \textit{lower bound} of a truly optimal \acs{QFI}.  

Moreover, \acs{ISS} is much more efficient than \acs{MOP}. Its exact time complexity depends on the strategy and it is hard to estimate. Our numerical data shows that for strategies that are similar to parallel or adaptive strategy and with the use of tensor networks the time complexity is quadratic in $N$. This is significantly better than exponential complexity of \acs{MOP} and allows to compute \acs{QFI} for significantly larger values of $N$ ($N\le5$ for \acs{MOP} vs $N\le 100$ for \acs{ISS}). Moreover, in \acs{ISS} we can control additional parameters like ancilla or bond dimension and consider strategies different from parallel and adaptive. Finally, while both methods can be used to get the optimal adaptive or parallel protocols, it is much more straightforward for \acs{ISS}, as it requires only inspecting the final values of the pre-QFI function optimized arguments.

\subsection{Upper bounds}\label{sec:bounds}
In Sec.~\ref{sec:mop} we explained that the \acs{QFI} can be computed via minimization of the norm of $\alpha$ \eqref{eq:alpha} over all Kraus representations $\{K_{\theta, k}\}$ of the channel $\Lambda_\theta$. It follows that in order to compute \acs{QFI} for parallel or adaptive strategy we need to optimize over Kraus representations $\{K^{(N)}_{\theta, k}\}$ of the whole channel $\Lambda^{(N)}_\theta=\Lambda^{\otimes N}_\theta$---a task that is exponentially hard in $N$. However, if we only want to obtain the upper bounds, we may dramatically reduce the complexity of the optimization, and reformulate the computation of the bound in a way that only minimization over Kraus representations of the \emph{single channel} $\Lambda_\theta$ is required  \cite{Demkowicz2012, Kurdzialek2023, kurdzialek2024bounds, Demkowicz2014, Kolodynski2013}.

In case of parallel strategies this leads to an upper bound of the form \cite{Escher2011, Demkowicz2012, Kolodynski2013}:
\begin{equation}\label{eq:bound_par}
    F^{(N)}_Q(\Lambda_\theta)  \leq  4N \min_{h} \left(\|\alpha(h)\|+(N-1)\|\beta(h)\|^2\right),
\end{equation}
where $\alpha(h)$ is defined in \eqref{eq:alpha}, while 
\begin{equation}
\beta(h) = \sum_k \dot K_{\theta, k}^\dagger(h) K_{\theta, k}(h).
\end{equation}
This bound allows immediate exclusion of the possibility of the Heisenberg scaling in the model---it is enough to show that there exist $h$ for which $\beta(h)=0$, which may be formulated as a simple algebraic condition \cite{Demkowicz2012, Zhou2021}.

In case of the adaptive strategy, one 
may again employ the idea of \acs{MOP} to bound the maximal increase of \acs{QFI} when the adaptive strategy is extended by one additional step including one more sensing channel \cite{Kurdzialek2023}:
\begin{multline}
    F_{\t{Q}}^{(N+1)}(\Lambda_\theta)\le F^{(N)}_{\t{Q}}(\Lambda_\theta) + \\ + 4\min_{h} \left[ \|\alpha(h)\| + \sqrt{F^{(N)}_{\t{Q}}(\Lambda_\theta)} \|\beta(h)\|\right].
\end{multline}
Both of these bounds can be computed efficiently regardless of the value of $N$ \cite{Kurdzialek2023}. They are implemented in the $\texttt{par\_bounds}$ and $\texttt{ad\_bounds}$ functions respectively.

Importantly, the bounds for parallel and adaptive strategies are asymptotically (for $N\rightarrow\infty$) equivalent and saturable \cite{Zhou2021, Kurdzialek2023}---there are strategies that asymptotically achieve the optimal value of \acs{QFI} predicted by the bounds.

Depending on whether there exist $h$ for which $\beta(h)=0$ we may have two possible asymptotic behaviors, namely 
the \emph{standard scaling} (\acs{SS}):
    \begin{equation}\label{eq:asym_ss}
        \lim_{N\rightarrow\infty}F^{(N)}_{Q}(\Lambda_\theta)/N = 4\min_{h,\beta(h)=0}\|\alpha(h)\|,
    \end{equation}
and the \emph{Heisenberg scaling} (\acs{HS}):
\begin{equation}\label{eq:asym_hs}
        \lim_{N\rightarrow\infty}F^{(N)}_{Q}(\Lambda_\theta)/N^2 = 4\min_h\|\beta(h)\|^2.
\end{equation}
In the package, the asymptotic bound can be computed using \texttt{asym\_scaling\_qfi}, a function that automatically also determines the character of the scaling.

Finally, there are methods to compute the bounds also in the case of correlated noise models that have been developed recently in \cite{kurdzialek2024bounds}. We do not provide the exact formulation of the optimization problem that one needs to solve to find the relevant bound as it is much more involved than in case of uncorrelated noise models discussed above. In the package, the correlated noise bounds may be computed using \texttt{ad\_bounds\_correlated} a finite number of channel uses, and \texttt{ad\_asym\_bound\_correlated} in order to obtain the asymptotic behavior. Unlike in the uncorrelated noise models, there is no guarantee here that the bounds are saturable. Still, they may be systematically tightened at the expense of increasing the numerical complexity, see Sec.~\ref{sec:correlated} for a more detailed discussion.

\section{Basic package usage---optimization of standard   strategies}\label{sec:basic}
In this section we present the basic ways to use the package that utilize high-level  functions from \texttt{protocols} and \texttt{bounds} subpackages. These functions allow the user to compute and bound the \acs{QFI} within three standard scenarios: single-channel, parallel and adaptive strategies, by simply specifying the parameter-encoding channel and the number of uses. 

In the following subsections we will apply these methods to study the problem of phase estimation in the presence of two paradigmatic decoherence models: dephasing (d) or amplitude damping (a).  

The parameter-encoding channels $\Lambda_\theta$ that we will consider are given by Kraus operators of the form
\begin{equation}
 K_{\theta,k} = U_{\theta} K_k, \quad U_\theta= e^{-\frac{\I}{2}\theta\sigma_z}, 
\end{equation}
where $\sigma_z$ is a Pauli $z$-matrix, $U_\theta$ represents 
unitary parameter encoding, whereas $K_k$ are 
Kraus operators corresponding to one of the two decoherence models. For simplicity, we will always assume that estimation is performed around the parameter value $\theta \approx 0$, so after differentiating the Kraus operators we set $\theta=0$---this does not affect the values of the \acs{QFI}
as finite unitary rotations could always be incorporated in control operations, measurements or the state preparation stage and hence will not affect the optimized value of \acs{QFI}.

In case of the \emph{dephasing model}, the two Kraus operators may be parametrized with a single parameter $p\in[0, 1]$:
\begin{equation}\label{eq:pardeph1}
        K^{(d)}_0=\sqrt{p}\bbone, \quad K^{(d)}_1 = \sqrt{1-p}\sigma_z,
\end{equation}
where $p=1$ corresponds to no dephasing, $p=1/2$ to the strongest dephasing case, where all equatorial qubit states are mapped to the maximally mixed state, while $p=0$ represents a phase flip channel. Equivalently, we may use a different Kraus representation where the two Kraus operators are given by:
\begin{equation}\label{eq:pardeph2}
        K^{(d)}_+=\frac{1}{\sqrt{2}} U_\epsilon, \quad K^{(d)}_-=\frac{1}{\sqrt{2}}  U_{-\epsilon},
    \end{equation}
which may be interpreted as random rotations by angles $\pm \epsilon$ with probability $1/2$ each---in order for the two representations to match we need to fix $p=\cos^2(\epsilon/2)$.

In case of \emph{amplitude} damping, the Kraus operators take the form
\begin{equation}\label{eq:paramp}
        K^{(a)}_0=\ket{0}\bra{0} + \sqrt{p} \ket{1}\bra{1},\, K^{(a)}_1 = \sqrt{1-p} \ket{0}\bra{1},
\end{equation}
where $p=1$ represents the no-decoherence case, while $p=0$ the full relaxation case, where all the states of the qubit are mapped to the ground $\ket{0}$ state.

Alternatively, dephasing and amplitude damping models can be considered in a continuous time regime. In this case, the state undergoes a process for some time $t$ which imprints on it information about the parameter $\omega$ related to $\theta$ by $\theta=\omega t$. The derivative of the state with respect to time is defined using the \emph{Lindbladian operator} $\mc L_\omega$:
\begin{equation}
    \frac{d\rho_{\omega, t}}{dt} = \mathcal{L}_{\omega}[\rho_{\omega, t}] = \frac{1}{i\hbar}[H_\omega, \rho_{\omega, t}] + \mc D[\rho_{\omega, t}],
\end{equation}
where $H_\omega = \hbar\omega\sigma_z/2$ is a Hamiltonian generating the evolution $U_\theta$ and $\mc D$ is a dissipation term. The dissipation term has a general form:
\begin{equation}
    \mc D[\rho] = \sum_i \gamma_i \left( L_i\rho L_i^\dagger - \frac{1}{2}\{L_i^\dagger L_i, \rho\} \right),
\end{equation}
where $\{\cdot,\cdot\}$ is the anticommutator, $L_i$ are jump operators that determine the type of dissipation and $\gamma_i\geq0$ are damping rates. In case of dephasing and amplitude damping only one pair $L_i, \gamma_i$ is required and it is respectively:
\begin{align}
    &L^{(d)} = \sigma_z/\sqrt{2},  &2p-1=e^{-\gamma^{(d)}t},\\
    &L^{(a)} = \ket{0}\bra{1}, &p=e^{-\gamma^{(a)}t}.
\end{align}

\subsection{Specifying the parameter-dependent channel}
\label{sec:specchannel}
First, let us start by explaining how to encode the channel using the \texttt{ParamChannel} class. We can do that in one of the four available ways:
\begin{itemize}
    \item{from a list of Kraus operators and their derivatives with respect to $\theta$},
    \item{from a \acs{CJ} matrix and its derivative over $\theta$,}
    \item{using a predefined creator function,}
    \item{from a Lindbladian, its derivative with respect to $\omega$ and a specified evolution time}.
\end{itemize}
The following listing shows how the above options are implemented in practice using the example of the dephasing channel \eqref{eq:pardeph1}:
\begin{lstlisting}[language=python]
import numpy as np
from qmetro import *

# Pauli z-matrix:
sz = np.array([[1, 0], [0, -1]])

# Kraus operators and their derivatives
# for dephasing channel:
p = 0.75
krauses = [
    np.sqrt(p) * np.identity(2),
    np.sqrt(1-p) * sz
]
dkrauses = [
    -1j/2 * sz @ K for K in krauses
]

# ParamChannel instance created from
# Kraus operators:
channel1 = ParamChannel(
    krauses=krauses, dkrauses=dkrauses
)

# CJ matrix and its derivative created
# from Kraus operators:
choi = choi_from_krauses(krauses)
dchoi = dchoi_from_krauses(
    krauses, dkrauses
)

# ParamChannel instance created from
# CJ matrix:
channel2 = ParamChannel(
    choi=choi, dchoi=dchoi
)

# ParamChannel instance created using
# creator function:
channel3 = par_dephasing(p)
\end{lstlisting}

By design the \texttt{ParamChannel} class represents a discrete time quantum evolution. Thus to create its objects from $\mc L_\omega, \dot{\mc{L}}_\omega$ one needs to move from continuous to discrete time regime by integrating over specified evolution time $t$. This can be done using \texttt{choi\_from\_lindblad} function from \texttt{qtools} which performs this integration for time-independent Lindbladian specified either as a function or a pair of Hamiltonian and a list of jump operators:
\begin{lstlisting}[language=python]
# continuing previous example

# Lindbladian and its derivative.
omega = 0.0 # Parameter.
t = 1 # Time.
# Dephasing strength:
gamma = -np.log(2*p-1)/t
# Hilbert space dimension
dim = 2

def lindblad(rho):
    # Rotation around z part:
    rot = 0.5j*omega * (rho@sz - sz@rho)
    # Dephasing part:
    deph = gamma * (sz@rho@sz - rho)
    return rot + deph

def dlindblad(rho):
    return 0.5j * (rho@sz - sz@rho)

choi, dchoi = choi_from_lindblad(
    lindblad, dlindblad, t, dim=dim
)
channel5 = ParamChannel(
    choi=choi, dchoi=dchoi
)

# or

# Hamiltonian
H = 0.5 * omega * sz
# Jump operators rescaled by sqrt(gamma)
Ls = [np.sqrt(gamma/2) * sz]
# Derivative of the Hamiltonian
dH = 0.5 * sz
# Derivative of the jump operators
dLs = [np.zeros_like(L) for L in Ls]
choi, dchoi = choi_from_lindblad(
    (H, Ls), (dH, dLs), t
)
channel6 = ParamChannel(
    choi=choi, dchoi=dchoi
)
\end{lstlisting}

Objects of the \texttt{ParamChannel} class can be used to compute $\rho_\theta = \Lambda_\theta(\rho)$ and $\dot\rho_\theta = \frac{d}{d\theta}\Lambda_\theta(\rho)$, e.g.:
\begin{lstlisting}[language=python]
channel = channel1
rho = np.array([
    [1, 0],
    [0, 0]
])

rho_t, drho_t = channel(rho)
\end{lstlisting}
or to obtain the \acs{CJ} operator from Kraus operators and vice versa, e.g.:
\begin{lstlisting}[language=python]
channel1 = ParamChannel(
    krauses=krauses, dkrauses=dkrauses
)
# CJ matrix:
choi = channel.choi()
# Derivative of CJ matrix:
dchoi = channel.dchoi()

channel2 = ParamChannel(
    choi=choi, dchoi=dchoi
)
# Kraus operators:
krauses = channel2.krauses()
# Kraus operators and their derivatives:
krauses, dkrauses = channel2.dkrauses()
\end{lstlisting}
They can be combined with each other to create new channels using:
\begin{itemize}
    \item scalar multiplication $\left(\alpha \Lambda_\theta \right)(\rho) = \alpha\Lambda_\theta(\rho)$, e.g.:
\begin{lstlisting}[language=python]
a = 0.2
new_channel = channel.scalar_mul(a)
# or equivalently
new_channel = a * channel
\end{lstlisting}
    \item addition $(\Lambda_\theta + \mr \Phi_\theta)(\rho) = \Lambda_\theta(\rho) + \Phi_\theta(\rho)$, e.g.:
\begin{lstlisting}[language=python]
new_channel = channel1.add(channel2)
# or equivalently
new_channel = channel1 + channel2
\end{lstlisting}
    \item composition $(\Lambda_\theta \circ \mr \Phi_\theta)(\rho) = \Lambda_\theta \left( \Phi_\theta(\rho) \right)$, e.g.:
\begin{lstlisting}[language=python]
new_channel = channel1.compose(
    channel2
)
# or equivalently
new_channel = channel1 @ channel2
\end{lstlisting}
    \item Kronecker product $\left( \Lambda_\theta \otimes \Phi_\theta \right)(\rho_1 \otimes \rho_2) = \Lambda_\theta(\rho_1) \otimes \Phi_\theta(\rho_2)$, e.g.:
\begin{lstlisting}[language=python]
new_channel = channel1.kron(channel2)
\end{lstlisting}
    \item Kronecker power $\Lambda_\theta^{\otimes N}  = \underbrace{\Lambda_\theta \otimes ... \otimes \Lambda_\theta}_{N}$, e.g.:
\begin{lstlisting}[language=python]
N = 3
new_channel = channel1.kron_pow(N)
\end{lstlisting}
\end{itemize}

\subsection{Single-channel QFI optimization}
\label{sec:single}
\begin{figure*}[t]
    \centering
    \includegraphics[width=0.9\textwidth]{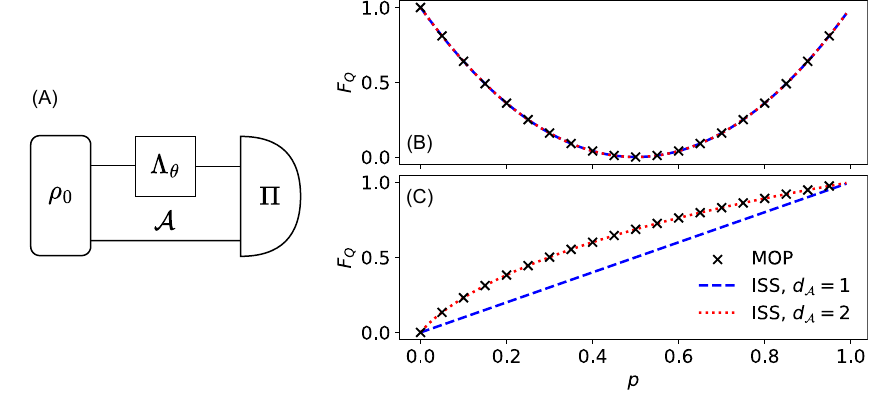}
    \caption{(A) Diagram of a strategy with a single parameter-dependent channel $\Lambda_\theta$ and ancilla $\mc A$. 
    Values of \acs{QFI} for (B) dephasing and (C) amplitude damping, for various values of $p$ and for different methods: \acs{MOP} - black $\times$, \acs{ISS} with $d_\mc{A}=1$ - blue dashed line, \acs{ISS} with $d_\mc{A}=2$ - red dotted line. 
    }
    \label{fig:single} 
\end{figure*}

After defining the parameter-dependent channel using the \texttt{ParamChannel} class, we may now easily compute the single-channel \acs{QFI} \eqref{eq:channelqfi}, including the possible use of an ancilla, see Fig.~\ref{fig:single}(A). 
The channel \acs{QFI}  can be computed using one of the two functions:
\begin{itemize}
    \item \texttt{iss\_channel\_qfi} for the \acs{ISS} method,
    \item \texttt{mop\_channel\_qfi} for the \acs{MOP} method.
\end{itemize}

The function \texttt{iss\_channel\_qfi} takes two arguments: the channel whose \acs{QFI} is computed and the dimension of the ancillary system. It returns five items: the optimized \acs{QFI}, a list of pre-QFI values per algorithm iteration, the density matrix of the optimal input state, the \acs{SLD} matrix, and the information whether the optimization converged successfully. 

On the other hand, \texttt{mop\_channel\_qfi} takes only the channel and returns only the \acs{QFI}---here, by the nature of the method, ancillary system dimension is unspecified, see Sec.~\ref{sec:mop}. The following code shows both of these functions in practice:
\begin{lstlisting}[language=python]
from qmetro import *

p = 0.75
channel = par_dephasing(p)

ancilla_dim = 2
iss_qfi, qfis, rho0, sld, status = iss_channel_qfi(channel, ancilla_dim)

mop_qfi = mop_channel_qfi(channel)
\end{lstlisting}
Fig.~\ref{fig:single}(B) and ~\ref{fig:single}(C) present plots of the results obtained using \texttt{iss\_channel\_qfi} and \texttt{mop\_channel\_qfi} for the two decoherence models respectively. Since the \acs{MOP} method gives the optimal \acs{QFI} implicitly assuming arbitrary ancilla dimension $d_\mc{A}$, these plots indicate that for the dephasing case (B) the ancilla is in fact unnecessary, while in the amplitude damping case (C) the ancillary system of dimension $d_\mc{A}=2$ is required for an optimal precision.

This example clearly shows the benefits of having two independent optimization procedures. One method (\acs{MOP}) provides the true optimal \acs{QFI}, but does not allow us to gain insight into the required ancilla size, while the other one (\acs{ISS})  allows us to study the impact of the ancilla size.


\subsection{Parallel strategy optimization}\label{sec:parallel}
\begin{figure*}[t]
    \centering
    \includegraphics[scale=1]{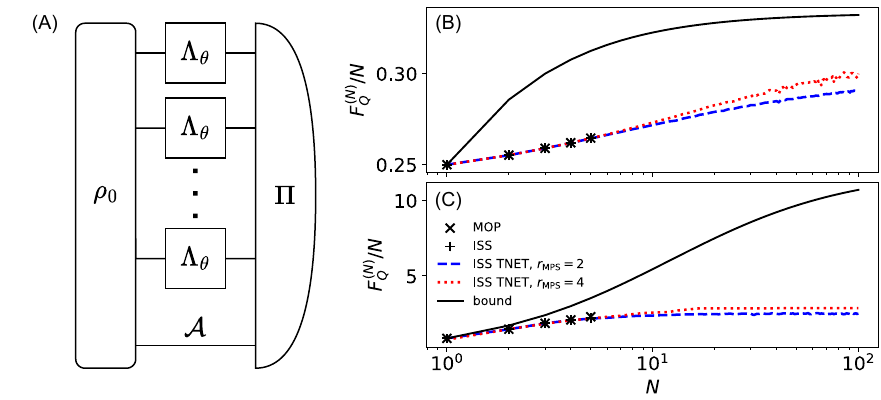}    
    \caption{(A) Diagram of a parallel strategy with multiple parametrized channels $\Lambda_\theta$ and ancilla $\mc A$.
    Values of \acs{QFI} normalized by the number of channel uses, $N$, for: (B)  dephasing ($p=0.75$), (C)  amplitude damping ($p=0.75$) and different methods: \acs{MOP} - black $\times$, simple \acs{ISS} with $d_\mc{A}=2$ - black +, tensor network \acs{ISS} with $d_\mc{A}=2, \; r_\mr{MPS} = \sqrt{r_\mf{L}}=2$ - blue dashed line, tensor network \acs{ISS} with $d_\mc{A}=2, \; r_\mr{MPS} = \sqrt{r_\mf{L}}=4$ - red dotted line, upper bound - black solid line. }
    \label{fig:parallel}
\end{figure*}

The parallel strategy refers to a setting in which $N$  copies of $\Lambda_\theta$ are probed in parallel by an entangled state and their output is collectively measured, see 
Fig.~\ref{fig:parallel}(A). It can also be understood as a strategy with a single channel $\Lambda_\theta^{(N)} = \Lambda_\theta^{\otimes N}$. The QMetro++ package provides several functions computing the \acs{QFI} for the parallel strategy. The simplest two, \texttt{iss\_parallel\_qfi} and \texttt{mop\_parallel\_qfi} are straightforward generalizations of \acs{ISS} and \acs{MOP} methods for parallel strategy:
\begin{lstlisting}[language=python]
from qmetro import *

p = 0.75
channel = par_dephasing(p)

N = 3
ancilla_dim = 2
iss_qfi, qfis, rho0, sld, status = iss_parallel_qfi(channel, N, ancilla_dim)

mop_qfi = mop_parallel_qfi(channel, N)
\end{lstlisting}

The downside of these two approaches is that they try to optimize \acs{QFI} over all possible states and measurements on an exponentially large Hilbert space. Hence, their time and memory complexity are exponential in the number of channel uses. In practice, this means that they can be used for relatively small $N$ ($N \lesssim  5$ in case of qubit channels).

This problem can be circumvented using tensor networks \acs{MPS} \eqref{eq:mps} and \acs{MPO} \eqref{eq:sld_mpo}. This approach is implemented in \texttt{iss\_tnet\_parallel\_qfi}, which requires two additional arguments specifying bond dimensions of the \acs{MPS} and the pre-SLD \acs{MPO}:
\begin{lstlisting}[language=python]
from qmetro import *

p = 0.75
channel = par_dephasing(p)

N = 3
ancilla_dim = 2
mps_bond_dim = 2
L_bond_dim = 4

qfi, qfis, psis, Ls, status = iss_tnet_parallel_qfi(channel, N, ancilla_dim, mps_bond_dim, L_bond_dim)
\end{lstlisting}
In contrast to the function implementing the basic \acs{ISS}, which returns the optimal input state and the \acs{SLD} in terms of arrays, this procedure gives them in the form of lists containing tensor components: \texttt{psis} = [$\psi_0$, $\psi_1$, ...] and \texttt{Ls} = [$\mf L_0$, $\mf L_1$, ...], see Eqs.~\eqref{eq:mps} and \eqref{eq:sld_mpo}.

Naturally, increasing ancilla and bond dimensions allows to achieve higher values of \acs{QFI} up to some optimal value, see Fig.~\ref{fig:parallel}(B) and (C). One might assess this value by running \texttt{iss\_tnet\_parallel\_qfi} for progressively larger dimension sizes, but it can also be upper-bounded using the \texttt{par\_bounds} function which gives a bound on the \acs{QFI} of the parallel strategy for all ancilla and bond dimension sizes (see Sec.~\ref{sec:bounds}):
\begin{lstlisting}[language=python]
from qmetro import *

p = 0.75
channel = par_dephasing(p)

N = 3

bound = par_bounds(channel, N)
\end{lstlisting}
The obtained value is an array \texttt{bound}=[$b_1, b_2, ..., b_N$] where $b_n$ is an upper bound on $F_Q^{(n)}(\Lambda_\theta)$. 

The numerical results are presented in Fig.~\ref{fig:parallel}(B) and (C). We considered the parallel strategy with ancilla dimension $d_\mc{A}$ and various bond dimensions $r_\mr{MPS}=\sqrt{r_\mf{L}}$. The relation between $r_{\rm{MPS}}$ and $r_{\mf{L}}$ was chosen based on the fact that the density matrix of the input state has bond dimension equal to $r_{\mr{MPS}}^2$, so taking $r_\mf{L}$ up to this value should not significantly affect the execution time (in practice, a smaller $r_\mf{L}$ is typically sufficient). Notice that for the dephasing case (B) results from \acs{MOP} and \acs{ISS} methods coincide perfectly and that \acs{QFI} for \acs{MPS} with $r_\mr{MPS}=4$ approaches the upper bound as $N$ increases. This suggests that in this case $d_\mc{A}=2$ is sufficient to saturate the bound. On the other hand, for the amplitude damping case (C) the results obtained using \acs{MOP} and \acs{ISS} differ, and already for $N=5$ the ancilla dimension $d_\mc{A}=2$ is suboptimal. Additionally, $F_Q^{(N)}(\Lambda_\theta)/N$ computed using \acs{MPS} quickly flattens out and the increase of bond dimension leads to a very small improvement. Therefore, we can conclude that in this case the dimensions $d_\mc{A}$ considered are insufficient to saturate the bound.

\subsection{Adaptive strategy optimization 
}\label{sec:adaptive}
\begin{figure*}[t]
    \centering
    \includegraphics[scale=1]{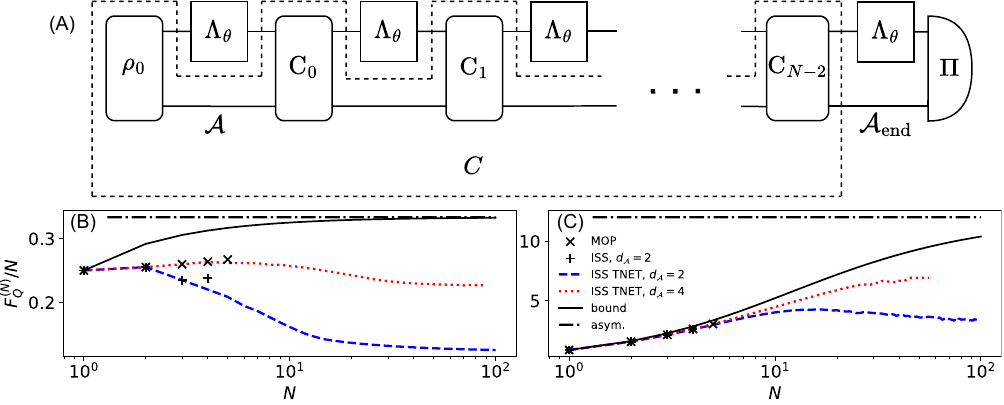}    
    \caption{(A) Diagram of an adaptive strategy with multiple parametrized channels $\Lambda_\theta$ and ancilla $\mc A$. When the optimization is performed over all quantum combs $C$ (area bounded by dashed lines) one can control the dimension of only the final ancilla ($\mc A_\mr{end}$). 
    Values of \acs{QFI} normalized by the number of channel uses, $N$, for: (B) dephasing ($p=0.75$) (C)  amplitude damping ($p=0.75$) and different methods: \acs{MOP} - black $\times$, simple \acs{ISS} with $d_\mc{A}=2$ - black +, tensor network \acs{ISS} with $d_\mc{A}=2$ - blue dashed line, tensor network \acs{ISS} with $d_\mc{A}=4$ - red dotted line, upper bound - black solid line, asymptotic bound - black dash-dotted line. Lack of data for $d_\mc{A}=4$, $N\ge 90$ (B) and $N \ge 60$ (C) is due to increasing numerical instability of \acs{ISS} with growing $N$.
    }
    \label{fig:adaptive}
\end{figure*}

In the adaptive strategy, we start with an initial state of the system and the ancilla, $\rho_0$, and we act on the system with $N$ parameter-encoding channels $\Lambda_\theta$, intertwined with control operations $\mr C_i$ which act both on the system and the ancilla,  see Fig.~\ref{fig:adaptive}(A). Equivalently, it is a strategy where $N$ parameter-encoding channels $\Lambda_\theta$ are plugged in between the teeth of a quantum comb $C$, see Fig.~\ref{fig:linkadaptive}, and then the resulting state is measured. In this strategy, one seeks to determine the value of \acs{QFI} for the optimal input state, measurement, and control operations (or equivalently the optimal comb and measurement). Notice that for sufficiently large ancilla dimension $d_\mc{A}$ and for $\mr C_i$ being appropriate SWAP gates one can simulate any parallel strategy with an adaptive strategy. Thus, assuming the ancilla is large enough, the \acs{QFI} for the adaptive strategy always upper-bounds the \acs{QFI} for the parallel strategy.

\acs{QFI} for adaptive strategy using the \acs{MOP} method can be computed by simply invoking \texttt{mop\_adaptive\_qfi}:
\begin{lstlisting}[language=python]
from qmetro import *

p = 0.75
channel = par_dephasing(p)

N = 3

qfi = mop_adaptive_qfi(channel, N)
\end{lstlisting}
This yields truly optimal \acs{QFI} corresponding to the optimal adaptive strategy.  

Alternatively, we may also use the \acs{ISS} method, 
with the help of \texttt{iss\_adaptive\_qfi} function. This procedure will perform the optimization over the whole comb as well.  Note that, the constraints on a quantum comb \eqref{eq:comb_cond} allow for its decomposition into control operations (\acs{CPTP} maps), but do not specify the ancilla dimension inside the comb. Therefore, in this case one can control only the dimension of the last ancilla which goes outside of the comb, indicated by the symbol $\mc A_\mr{end}$ in Fig.~\ref{fig:adaptive}(A):
\begin{lstlisting}[language=python]
from qmetro import *

p = 0.75
channel = par_dephasing(p)

N = 3
# Dimension of the last ancilla:
ancilla_dim = 2

qfi, qfis, comb, sld, status = iss_adaptive_qfi(channel, N, ancilla_dim)
\end{lstlisting}
Apart from the standard outputs for the \acs{ISS}-type functions, \texttt{iss\_adaptive\_qfi}  returns \texttt{comb} which is a \acs{CJ} matrix of the optimal comb.

Finally, if we want to make use of the tensor-network structure, and perform the optimization over separate control operations, we can use  \texttt{iss\_tnet\_adaptive\_qfi}. Notice that in this approach we control ancilla dimension at each step and thus, in general, the obtained \acs{QFI} will be smaller than the one we got with \texttt{iss\_adaptive\_qfi}---the advantage here is that the optimization will be much more efficient (not to mention that this is also the only way to go to the larger $N$ limit):
\begin{lstlisting}[language=python]
from qmetro import *

p = 0.75
channel = par_dephasing(p)

N = 3
# Dimension of the ancilla:
ancilla_dim = 2

qfi, qfis, teeth, sld, status = iss_adaptive_qfi(channel, N, ancilla_dim)
# teeth can be decomposed into rho0 and
# a list of controls:
rho0, *Cs = teeth
\end{lstlisting}
The returned argument \texttt{teeth} represents the comb's teeth and it is a list consisting of the optimal initial state and \acs{CJ} matrices of the optimal control operations \texttt{teeth} $= [\rho_0, C_0, C_1, ..., C_{N-2}]$.

The upper bound for the adaptive strategy is computed using \texttt{ad\_bounds} function:
\begin{lstlisting}[language=python]
from qmetro import *

p = 0.75
channel = par_dephasing(p)

N = 3
bound = ad_bounds(channel, N)
\end{lstlisting}
Additionally, one can compute the value of the bound in the limit $N\rightarrow \infty$ which is asymptotically saturable and the same for both the parallel and the adaptive strategy, see Sec.~\ref{sec:bounds}:
\begin{lstlisting}[language=python]
from qmetro import *

p = 0.75
channel = par_dephasing(p)

c, k = asym_scaling_qfi(channel)
\end{lstlisting}
where the returned values should be interpreted in the following way $c=\lim_{N\rightarrow\infty}F_{Q}^{(N)}(\Lambda_\theta)/N^{k}$.

Values of \acs{QFI}, bounds, and asymptotic \acs{QFI} for the adaptive strategy are presented in Figs.~\ref{fig:adaptive} (B) and (C). In contrast to the parallel strategy, this time the amplitude damping case (C) shows results which are much closer to the bound. 

Notice that there is no data for amplitude damping and ancilla dim $d_\mc{A}=4$ and $N \ge 60$. This is due to the fact that for this noise model, the \acs{QFI} attains a large value and simultaneously optimal control operations $\mr C_i$ take the state close to the pure state. These two processes combined create objects (tensors, matrices etc.) which hold values from a very wide range at the same time. This causes loss of precision and then numerical instability. A similar issue appears for dephasing but for much larger $N\simeq90$.

\subsection{Correlated noise models}
\label{sec:correlated}
\begin{figure*}[t]
    \centering
    \includegraphics[scale=1]{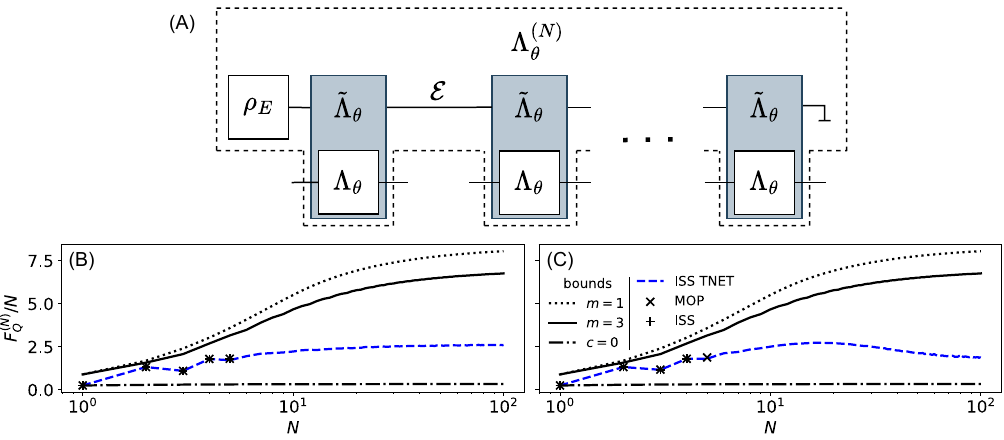}
    \caption{(A) Diagram of $N$ parametrized channels $\Lambda_\theta$ with noise correlations. Extended channels $\tilde\Lambda_\theta$ (gray) act on the system as $\Lambda_\theta$ and are connected with each other by the environment space $\mc E$. The environment system $\mc E$ starts in the state $\rho_E$ and is not accessed (is traced out) at the end. 
    Values of \acs{QFI} normalized by the number of channel uses, $N$, for anti-correlated ($c=-0.75$) dephasing ($p=0.75$), different strategies: (B) parallel, (C) adaptive and various methods: \acs{MOP} - black $\times$, simple \acs{ISS} with $d_\mc{A}=2$ - black +, tensor network \acs{ISS} with $d_\mc{A}=2$ (parallel: $r_\mathrm{MPS}=\sqrt{r_\mf{L}}=2$) - blue dashed line, upper bound with block size $m=1$ - black dotted line, upper bound with block size $m=3$ - black solid line and upper bound for uncorrelated channels $c=0$ - black dash-dotted line. Notice that since there are no separate bounds for parallel strategy with correlated channels the upper bounds (black dotted and solid lines) are identical in (B) and (C). 
    }
    \label{fig:corr}
\end{figure*}
The QMetro++ also provides a simple way to compute the \acs{QFI} for the adaptive and the parallel strategies in scenarios when parameter-encoding channels act on an additional environment space $\mc E$ which directly connects subsequent channels, see Fig.~\ref{fig:corr}(A). This approach allows in particular to model noise correlations affecting the sensing channels.

To illustrate this, let us consider a Kraus representation for the dephasing model with Kraus operators $K_\pm$ acting as Bloch-sphere rotations by $\pm\epsilon$ around $z$ axis, see Eq.~\eqref{eq:pardeph2}. We want the signs of these rotations to form a binary Markovian chain. The conditional probability of a rotation with sign $s_i\in\{+,-\}$ in channel $i$ given $s_{i-1}$ in channel $i-1$ is:
\begin{equation}\label{eq:corr_prob}
    T_{i|i-1}(s_i|s_{i-1}) = \frac{1}{2}(1+s_i s_{i-1} c),
\end{equation}
where $c\in[-1, 1]$ is a correlation parameter with $c=0$ meaning no correlations, $c=1$ maximal positive correlations and $c=-1$ maximal negative correlations. The initial probabilities are $p(\pm)=1/2$. 

\begin{figure}[t]
\includegraphics[scale=0.75]{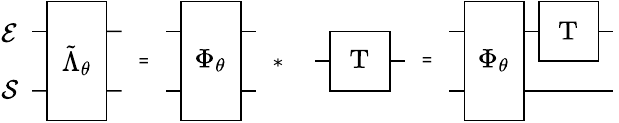}.
\caption{Modeling of correlated dephasing process, by extending the action of the channel on a two-dimensional environment space.}
\label{fig:extchannel}
\end{figure}

This process can be modeled with the help of a two-dimensional environment Hilbert space $\mc E$ which will carry the information about the sign of the rotation in the preceding step. 
Let us introduce a new channel acting on both the environment and the system $\tilde\Lambda_\theta = \Phi_\theta*\mr T$, see  Fig.~\ref{fig:extchannel}, where:
\begin{multline}\label{eq:corr_constr}
    \Phi_\theta : \mc L(\mc E \otimes \mc S) \ni \rho \mapsto V_\theta\rho V_\theta^\dagger\in \mc L(\mc E \otimes \mc S), \\ V_\theta=\sqrt{2}\sum_{s=\pm} \ket{s}\bra{s} \otimes K_{s, \theta},
\end{multline}
and
\begin{multline}
\label{eq:corr_constr2}
    \mr T : \mc L(\mc E) \ni \rho_E\mapsto \sum_{s, r = \pm} T_{sr}\rho_E T_{sr}^\dagger\in \mc L(\mc E ), \\ 
    T_{sr}=\sqrt{\frac{1+s r c}{2}}\ket{s}\bra{r}.
\end{multline}
Clearly, $\tilde\Lambda_\theta$ acts as $\Lambda_\theta$ on the system---$\Tr_{\mc E} \tilde\Lambda_\theta(\rho)=\Lambda_\theta(\Tr_{\mc E} \rho)$ for any state $\rho$. The unitary channel $\Phi_\theta$ acts with $K_\pm$ on the system, depending on the state of the environment being $\ket \pm$, and it stores the information about the occurrence of $K_\pm$ in the state $\ket\pm$ accordingly. Then $\mr T$ modifies the state of $\mc E$ according to probabilities \eqref{eq:corr_prob}. The initial probabilities are encoded in the initial environment state $\rho_E= p(+)\ket{+}\bra{+}+p(-)\ket{-}\bra{-}=\bbone_2/2$.

Channel $\tilde\Lambda_\theta$ can be easily constructed using the \texttt{link\_env} method which links two channels through their common  environment:
\begin{lstlisting}[language=python]
from numpy import array, kron, sqrt
from qmetro import (
    par_dephasing, ket_bra, ParamChannel
)

p = 0.75
c = -0.75

# rot_like=True gives dephasing in
# the second form:
Lambda = par_dephasing(p, rot_like=True)
krauses, dkrauses = Lambda.dkrauses()
# Rotations by plus and minus epsilon:
Kp, Km = krauses
# Derivatives of Krauses:
dKp, dKm = dkrauses

# Plus state:
plus = array([1, 1]) / sqrt(2)
# Minus state:
minus = array([1, -1]) / sqrt(2)

# V_theta matrix:
V = kron(ket_bra(plus, plus), Kp)
V = V + kron(ket_bra(minus, minus), Km)
V = sqrt(2) * V

# Derivative of V_theta matrix:
dV = kron(ket_bra(plus, plus), dKp) 
dV += kron(ket_bra(minus, minus), dKm)
dV = sqrt(2) * dV

# env_dim=d tells ParamChannel
# constructor that the channel acts
# on the d-dimensional environment space
# and the system
Phi = ParamChannel(
    krauses=[V], dkrauses=[dV],
    env_dim=2
)

# T channel:
T_krauses = []
for s in (1, -1):
    for r in (1, -1):
        x = sqrt((1+s*r*c)/2)
        Tsr = x * ket_bra(
            array([1, s]), array([1, r])
        )/2
        T_krauses.append(Tsr)
# If not provided, the derivative is
# zero by default:
T = ParamChannel(
    krauses=T_krauses, env_dim=2
)

Lambda_tilde = Phi.link_env(T)
# or equivalently:
Lambda_tilde = Phi * T
\end{lstlisting}
When provided with \texttt{env\_dim} argument, the constructor of \texttt{ParamChannel} will assume that the channel acts on a space $\mc E \otimes \mc S$ with $\dim\mc E$ given by the argument. All functions implementing \acs{MOP} and \acs{ISS} algorithms from Sec.~\ref{sec:parallel} and Sec.~\ref{sec:adaptive} can take such a channel as their argument and optimize the corresponding \acs{QFI} for a sequence of channels as in Fig.~\ref{fig:corr}(A):
\begin{lstlisting}[language=python]
...

from numpy import identity
from qmetro import iss_tnet_adaptive_qfi

N = 3
d_a = 2
# QFI for the adaptive strategy and
# correlated noise:
qfi, *_ = iss_tnet_adaptive_qfi(
    Lambda_tilde, N, d_a
)

# Input state of the environment is by
# default a maximally mixed state but
# can be specified using env_inp_state
# parameter:
rhoE = identity(2) / 2
qfi, *_ = iss_tnet_adaptive_qfi(
    Lambda_tilde, N, d_a,
    env_inp_state=rhoE
)
\end{lstlisting}
Bounds in this case can be computed using \texttt{ad\_bounds\_correlated} and \texttt{ad\_asym\_bound\_correlated}. In contrast to the bounds for the uncorrelated case these bounds are not necessarily tight asymptotically. They are computed by upper-bounding \acs{QFI} for adaptive strategy where every $m$-th control operation has access to the environment, for some integer $m$. This leads to a leakage of information from the environment, which weakens the tightness of the bound. 
Naturally, larger value of $m$ makes the bound tighter but at the same time it exponentially increases the execution time. Realistically, one can compute the bound for $m \lesssim 5$ in case of qubit channels. An exemplary code to compute the bound is given below:
\begin{lstlisting}[language=python]
...

from qmetro import ad_bounds_correlated

N = 10
m = 3
ns, bound = ad_bounds_correlated(
    Lambda_tilde, N, m
)
\end{lstlisting}
where \texttt{ns} $=[1, 1+m, 1+2m, ..., ]$, \texttt{bound}$=[b_1, b_{1+m}, b_{1+2m}, ...]$ and $b_n$ is an upper bound on $F_{Q}^{(n)}(\Lambda_\theta)$.

Note that the construction \eqref{eq:corr_constr}, \eqref{eq:corr_constr2} is not unique and correlations \eqref{eq:corr_prob} can be modeled using different $\tilde\Lambda_\theta$. For example a quantum channel:
\begin{multline}\label{eq:corr_constr3}
    \mr t : \mc L(\mc E) \ni \rho \mapsto \sum_{s, r} t_{sr}\rho\, t_{sr}^\dagger\in \mc L(\mc E ), \\ 
    t_{sr}=\sqrt{\frac{1+s r\sqrt c}{2}}\ket{s}\bra{r},
\end{multline}
satisfies $\mr t\circ \mr t=\mr T$ thus $\tilde\Lambda_\theta = \mr t *\Phi_\theta*\mr t$ will give the same $\Lambda^{(N)}_\theta=\tilde\Lambda_\theta*...*\tilde\Lambda_\theta$ as before. This observation is important for calculation of upper-bounds because of the aforementioned fact that they are not computed for the whole $\tilde\Lambda_\theta^{(N)}$ but for $N/m$ blocks of size $m$ and the construction influences how these blocks begin and end. For example, it turns out that the original construction \eqref{eq:corr_constr2} gives in this case an unrealistic \acs{HS} and the optimal bound is achieved for \eqref{eq:corr_constr3}, see \cite{kurdzialek2024bounds} for more discussion.

Values of the \acs{QFI} for anti-correlated ($c=-0.75$) dephasing case are shown in Figs.\ref{fig:corr} (B) and (C). Note, that the presence of (anti)correlations significantly increases the obtained \acs{QFI} which becomes even larger then the upper-bounds for uncorrelated case. Interestingly, we can see that the adaptive strategy with $d_\mc{A}=2$ is optimal for $N\le 5$, unlike in the uncorrelated dephasing case, see Fig.~\ref{fig:adaptive}(B).

\section{Advanced package usage---optimization of strategies with arbitrary structures}
\label{sec:advanced}

\begin{figure*}[t!]
    \centering
    \includegraphics[scale=1]{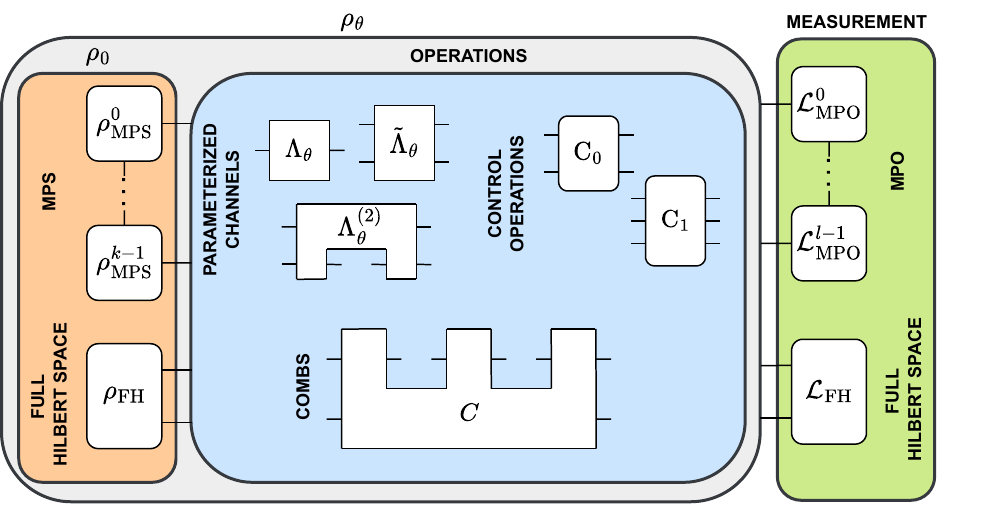}
    \caption{Diagram of a strategy with arbitrary structure that can be optimized using \texttt{iss\_opt}. The input state, $\rho_0$, might consist of \acs{MPS}s and/or states on a full Hilbert space. The input state goes through an arbitrary arrangement of parametrized channels, control operations and combs. The requirement is that each operation’s output can be connected only to the input of a different operation such  that there are no causal cycles. Finally, the state obtained after application of each operation, $\rho_\theta$, is measured. The measurement is specified by the form of the pre-SLD matrix $\mf L$ which can be an \acs{MPO} and/or a matrix on a full Hilbert space. States, control operations, and combs can be marked as variable or constant, the pre-SLD must be variable and the parametrized channels must be constant.}
    \label{fig:gen_str}
\end{figure*}

In addition to the functions presented in Sec.~\ref{sec:basic}, the QMetro++ allows the user to define their own strategy and then optimize it using the \acs{ISS} algorithm. Strategies are defined via a straightforward-to-use symbolic programming framework. The user creates a tensor network representing their strategy where some of the nodes are marked as constants and others are marked as variables to be optimized over---let us call the set of variables $\mc V$. This strategy is then plugged into the function \texttt{iss\_opt} which, based on the provided data, constructs the pre-QFI:
\begin{equation}\label{eq:preqfi_many}
    F(\mc V) = 2\Tr(\dot\rho_\theta\mf L) - \Tr(\rho_\theta\mf L^2),
\end{equation}
where $\rho_\theta$ and $\mf L$ are defined as in Fig.~\ref{fig:gen_str} and $\dot\rho_\theta$ is created from $\rho_\theta$ using the Leibniz (chain) rule. Then the function optimizes the above pre-QFI over each variable node one by one in a typical \acs{ISS} manner.

In Sec.~\ref{sec:qmtensor} we explain the classes of QMetro++ tensors (placed in \texttt{qmtensor} folder, see Fig.~\ref{fig:scheme}) which are the building blocks of strategies. Then in Sec.~\ref{sec:collisional} we show how such a strategy can be created using the example of the collisional metrological model.

\subsection{QMetro++ tensors}\label{sec:qmtensor}

A general strategy in QMetro++ is created by defining it as a tensor network of:
\begin{itemize}
    \item density matrices,
    \item elements of \acs{MPO}s of density matrices, in particular density matrices of \acs{MPS}s (see \eqref{eq:mps_density_mat}), 
    \item parameter-encoding channels, 
    \item control operations,
    \item quantum combs,
    \item measurements,
    \item elements of measurement \acs{MPO}s.
\end{itemize}
connected by their physical and bond spaces, as in the diagram in Fig.~\ref{fig:gen_str}.
Some input states, control operations, and quantum combs as well as all measurements, are marked as variables of the pre-QFI function \eqref{eq:preqfi_many} and they will be optimized by the \acs{ISS} algorithm. Each non-measurement node is represented using its tensor representation (see a detailed explanation in the Appendix~\ref{sec:tensor_networks}) and measurements are represented via a tensor representation of the pre-SLD matrix $\mf L$.
\begin{figure*}[t]
    \centering
    \includegraphics[width=\linewidth]{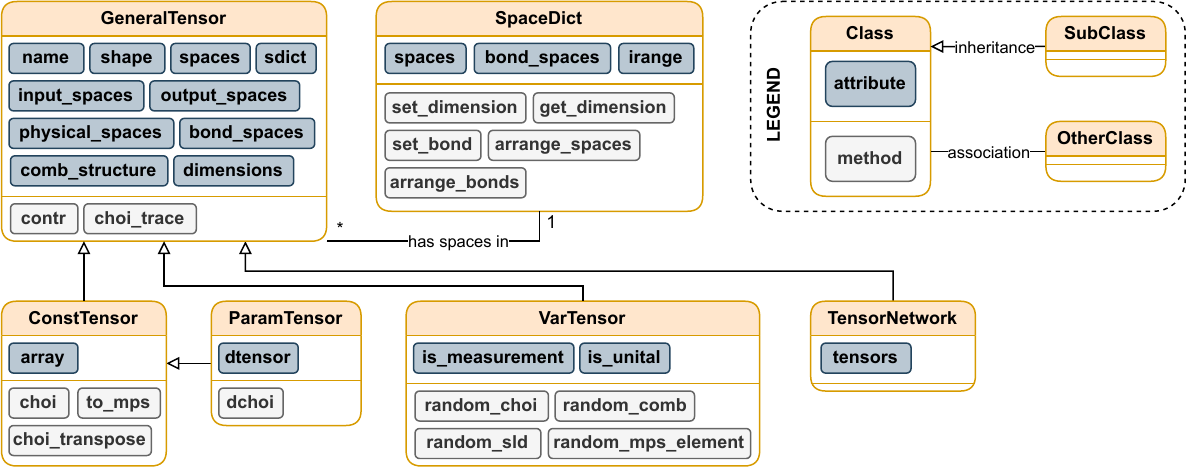}
    \caption{\acs{UML} diagram of classes used for creation of strategies with their most important attributes and methods. Subclasses inherit all attributes and methods of their parent class. Instances of \texttt{GeneralTensor} and its subclasses can access a \texttt{SpaceDict} instance through their \texttt{sdict} attribute. Two tensors can be contracted with each other only when they have the same \texttt{sdict} attribute.}
    \label{fig:classes}
\end{figure*}

In practice, this means that each node of the network has to be initialized as an object of a QMetro++ tensor class (see Fig.~\ref{fig:classes}). There are five kinds of tensor classes:
\begin{itemize}
    \item \texttt{GeneralTensor} - a class containing attributes and methods common to all tensors, such as the shape or the contraction operation. It is an \emph{abstract class} \cite{bloch2001} i.e. it is only a blueprint for other classes to inherit from and it is not meant to be used in any computations directly.
    \item \texttt{ConstTensor} - a class representing tensors that stay constant during the optimization process.
    \item \texttt{ParamTensor} - a subclass of \texttt{ConstTensor} which has an additional attribute defining its derivative over the measured parameter $\theta$. It is meant to represent parameter-encoding channels $\Lambda_\theta$.
    \item \texttt{VarTensor} - a class representing tensors that are optimized by the \acs{ISS} algorithm. Members of this class store information specifying how they should be initialized and optimized (for example that they are \acs{CPTP} channels, quantum combs or pre-SLD matrices) but they do not have any value---they are only symbols.
    \item \texttt{TensorNetwork} - networks of constant, parametrized and variable tensors.
\end{itemize}
Additionally, there is one more class, \texttt{SpaceDict}, that stores the information about the names, dimensions and types of the respective spaces. For example:
\begin{lstlisting}[language=python]
import numpy as np
from qmetro import *

sd = SpaceDict()
# Adding physical spaces with names
# 'a', 'b' and dimension 2 to sd.
sd['a'] = 2
sd['b'] = 2

# Calling sd['a'] will return
# the dimension of space 'a'.
print(sd['a']) # Prints 2.

# Adding bond space 'x':
sd.set_bond('x', 4)
print(sd['x']) # Prints 4.

# Accessing range of appropriate index:
print(sd.irange['a']) # Prints 4.
print(sd.irange['x']) # Prints 4.
\end{lstlisting}
Note that the range of the physical index is equal to the square of the dimension of the corresponding space. This is because physical indices are created by joining two indices of a \acs{CJ} operator. However, this is not the case for bond indices which are not related to any Hilbert spaces. See detailed discussion in Appendix~\ref{sec:tensor_networks}).

Instances of tensors are created on spaces defined in the \texttt{SpaceDict}:
\begin{lstlisting}[language=python]
...

# Constant tensor can be initialized
# from a CJ matrix:
id_ab = np.identity(sd['a']*sd['b'])
ct0 = ConstTensor(
    ['a', 'b'], choi=id_ab, sdict=sd
)

# Its tensor representation is in array
# attribute:
ct0_ten_rep = ct0.array 
# Order of its indices is in spaces
# attribute:
spaces = ct0.spaces

# Constant tensor can be initialized
# from its tensor representation
ct1 = ConstTensor(
    spaces, array=ct0_ten_rep, sdict=sd
)

# Parametrized tensor requires both its 
# value and derivative. For example
# a tensor with trivial derivative will
# be:
zeros = np.zeros_like(id_ab)
pt = ParamTensor(
    spaces, sdict=sd,
    choi=id_ab, dchoi=zeros
)
\end{lstlisting}

Variable tensors require only spaces. Their type is derived from additional attributes:
\begin{lstlisting}[language=python]
...

spaces = ['a', 'b']
bonds = ['x']

# Variable tensor:
vt0 = VarTensor(spaces, sd)

# Types:
# > state:
vt1 = VarTensor(
    spaces, sd, output_spaces=spaces
)
# > element of a density matrix of
# an MPS:
vt1 = VarTensor(
    spaces + bonds, sd,
    output_spaces=spaces
)
# > CPTP map of a channel 'a'->'b':
vt1 = VarTensor(
    spaces, sd, output_spaces=['b']
)
# > pre-SLD:
vt2 = VarTensor(
    spaces, sd, is_measurement=True
)
# > element of a pre-SLD MPO:
vt2 = VarTensor(
    spaces + bonds, sd,
    is_measurement=True
)
\end{lstlisting}

Tensors can be contracted using either \texttt{contr} method, function with the same name or the star `*' symbol:
\begin{lstlisting}[language=python]
from qmetro import *

channel = par_dephasing(0.75)
choi = channel.choi()

sd = SpaceDict()
spaces = ['a', 'b', 'c']
for space in spaces:
    sd[space] = 2

ct0 = ConstTensor(['a', 'b'], choi, sd)
ct1 = ConstTensor(['b', 'c'], choi, sd)

# Contraction of 'b' space.
ct2 = ct0.contr(ct1)
# or equivalently 
ct2 = contr(ct0, ct1)
# or equivalently
ct2 = ct0 * ct1
\end{lstlisting}
While contraction of two \texttt{ConstTensor}s  simply creates a new \texttt{ConstTensor} this is not the case when one of the contracted tensors is a \texttt{VarTensor} because it does not have any value. Contraction with a \texttt{VarTensor}  creates a \texttt{TensorNetwork} consisting of the two tensors:
\begin{lstlisting}[language=python]
...

ct = ConstTensor(['a', 'b'], choi, sd)
vt = VarTensor(['b', 'c'], sd)
# Tensor network of ct and vt connected
# by 'b' space:
tn = ct * vt 

# Elements of tensor network can be 
# accessed using tensor and name
# attributes:
print(tn.tensors[ct.name] is ct) # True.
print(tn.tensors[vt.name] is vt) # True.
\end{lstlisting}
Similarly, a contraction with a 
\texttt{TensorNetwork} creates another \texttt{TensorNetwork}. The results of all possible contraction combinations are presented in Table~\ref{tab:link_types}. 
\begin{table}[t]
    \centering
    \begin{tabular}{||c|c|c|c|c|c||}
        \hline\hline
         * & GT & CT & PT & VT & TN   \\\hline
         GT & Error & Error & Error & Error & Error   \\\hline
         CT & Error & CT & PT & TN & TN   \\\hline
         PT & Error & PT & PT & TN & TN   \\\hline
         VT & Error & TN & TN & TN & TN   \\\hline
         TN & Error & TN & TN & TN & TN   \\\hline\hline
    \end{tabular}
    \caption{Resulting type of a contraction of various tensor pairs: GT--\texttt{GeneralTensor}, CT--\texttt{ConstTensor}, PT--\texttt{ParamTensor}, VT-- \texttt{VarTensor}, TN--\texttt{TensorNetwork}. Contraction with generalized tensor raises an error because instances of this class are not meant to be used in calculations. When \texttt{ParamTensor} is contracted with a \texttt{ConstTensor} or a \texttt{ParamTensor} the derivative of the result is computed using the Leibniz (chain) rule.
    }
    \label{tab:link_types}
\end{table}

\subsection{Collisional strategy}\label{sec:collisional}

\begin{figure*}[t]
    \centering
    \includegraphics[scale=1]{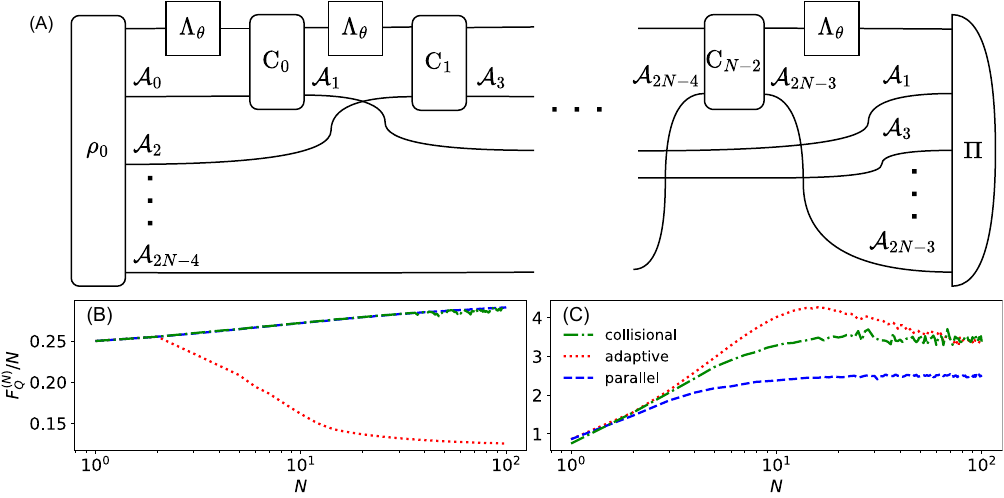}
    \caption{(A) Diagram of a collisional strategy with multiple parametrized channels $\Lambda_\theta$ and ancillas $\mc A_i$.
    Values of \acs{QFI} normalized by the number of channel uses, $N$, for: (B) dephasing ($p=0.75$), (C)  amplitude damping ($p=0.75$) and different strategies: adaptive ($d_\mc{A}=2$) - blue dashed line, parallel ($d_\mc{A}=2,\; r_\mathrm{MPS}=\sqrt{r_\mf{L}}=2$) - red dotted line, collisional ($d_\mc{A}=2,\; r_\mathrm{MPS}=\sqrt{r_\mf{L}}=2$) - green dash-dotted line. Fluctuations for larger $N$ are due to numerical instabilities.
    }
    \label{fig:passive}
\end{figure*}

We will describe the creation of a custom strategy using the example of a \emph{collisional strategy} depicted in Fig.~\ref{fig:passive}(A). Physically, we can imagine this scheme to represent a situation where multiple particles, potentially entangled with each other initially, approach the sensing system one by one. They interact locally with the system via control operations $\mr C_i$, carrying away information about the parameter while at the same time causing some back-action on the sensing system itself. 

This strategy may be viewed as a hybrid of the parallel and the adaptive strategies where each control operation $\mr C_i$ has its own ancilla which is measured immediately after the action of $\mr C_i$. Notice that if all $\mr C_i$ were set to SWAP gates one would regain the parallel strategy. On the other hand, this strategy does not allow control operations to communicate with each other like in the most general adaptive strategy. Therefore the \acs{QFI} for this strategy must be at least as good as for the parallel strategy but in general will be worse than for the adaptive strategy with an ancilla dimension $d_\mc{A}^{N-1}$, where $d_\mc{A}$ is the dimension of a single ancillary system in the collisional strategy.

We can now create a tensor network representing the collisional strategy. In the following we will use functions from the \texttt{qmtensor/creators} folder, see Fig.~\ref{fig:scheme}. These functions greatly simplify the process by creating entire structures (for example the density matrix of an \acs{MPS}) and they make the code more readable. Let us start by defining the necessary spaces:
\begin{lstlisting}[language=python]
from qmetro import *

# Number of parametrized channels:
N = 10
d_a = 2 # Ancilla dimension.

sd = SpaceDict()

# arrange_spaces will create a series of
# spaces ('INP', 0), ..., ('INP', N-1)
# with specified dimension and it will
# return a list of their names:
# > input spaces of parametrized
#   channels:
inp = sd.arrange_spaces(N, 2, 'INP')
# > output spaces of parametrized
#   channels:
out = sd.arrange_spaces(N, 2, 'OUT')
# > ancillas:
anc = sd.arrange_spaces(2*N-2, 2, 'ANC')
\end{lstlisting}
The input state can be created using \texttt{input\_state\_var} or \texttt{mps\_var\_tnet}: 
\begin{lstlisting}[language=python]
# First input space and ancillas with
# even indices:
spaces = [inp[0]] + anc[::2]
rho = input_state_var(
    spaces, name='RHO0', sdict=sd
)

# In case we want to optimize over MPSs:
r_mps = 2 # MPS bond dimension.
rho0 = input_state_var(
    spaces, 'RHO0', sd,
    mps_bond_dim=r_mps
)
# or equivalently:
rho0, rho0_names, rho0_bonds = mps_var_tnet(spaces, 'RHO0', sd, r_mps)
\end{lstlisting}
Both of these functions create the same \acs{MPS} but \texttt{mps\_var\_tnet} returns additionally names of its components and bond spaces which might become useful later for accessing the optimization results.

Tensors of parameter-encoding channels can be created using \texttt{tensor} method from the \texttt{ParamChannel} class:
\begin{lstlisting}[language=python]
# Parameter-encoding channel:
channel = par_dephasing(0.75)

# This will be the tensor network of our
# strategy:
tnet = rho0

# Adding channels to tnet:
for i in range(N):
    inp_i = inp[i] # i-th input space.
    out_i = out[i] # i-th output space.
    pt = channel.tensor(
        [inp_i], [out_i], sdict=sd,
        name=f'CHANNEL{i}'
    )
    tnet = tnet * pt
\end{lstlisting}
We want the control operations to be \acs{CPTP} maps, hence we will use \texttt{cptp\_var} creator:
\begin{lstlisting}[language=python]
for i in range(N-1):
    # i-th control input spaces:
    c_inp = [out[i], anc[2*i]]
    # i-th control output spaces:
    c_out = [inp[i+1], anc[2*i+1]]
    # Variable control operation:
    vt = cptp_var(
        c_inp, c_out, sdict=sd,
        name=f'CONTROL{i}'
    )
    tnet = tnet * vt
\end{lstlisting}
Similarly to the input state, the measurement can be created in one of two ways:
\begin{lstlisting}[language=python]
# Ancillas with odd indices and last
# output space:
spaces = anc[1::2] + [out[-1]]
Pi = measure_var(
    spaces, name='Pi', sdict=sd
)

# In case we want to optimize over
# pre-SLD which is MPO:
r_L = r_mps*r_mps # pre-SLD bond dim.
Pi = measure_var(
    spaces, 'Pi', sd, bond_dim=r_L
)
# or equivalently:
Pi, Pi_names, Pi_bonds = mpo_measure_var_tnet(spaces, 'Pi', sd, r_L)

tnet = tnet * Pi
\end{lstlisting}
In summary, the code is:
\begin{lstlisting}[language=python]
from qmetro import *

ch = par_dephasing(0.75) # Channel.
N = 10 # Number of channels.
d_a = 2 # Ancilla dimension.
r_mps = 2 # MPS bond dimension.
r_L = r_mps**2 # pre-SLD bond dimension.

sd = SpaceDict()
inp = sd.arrange_spaces(N, 2, 'INP')
out = sd.arrange_spaces(N, 2, 'OUT')
anc = sd.arrange_spaces(2*N-2, 2, 'ANC')

rho0_spaces = [inp[0]] + anc[::2]
rho0, rho0_names, rho0_bonds = mps_var_tnet(rho0_spaces, 'RHO0', sd, r_mps)

tnet = rho0
for i in range(N):
    pt = ch.tensor(
        [inp[i]], [out[i]], sdict=sd,
        name=f'CHANNEL{i}'
    )
    tnet = tnet * pt

for i in range(N-1):
    vt = cptp_var(
        [out[i], anc[2*i]],
        [inp[i+1], anc[2*i+1]],
        sdict=sd, name=f'CONTROL{i}'
    )
    tnet = tnet * vt

spaces = anc[1::2] + [out[-1]]
Pi, Pi_names, Pi_bonds = mpo_measure_var_tnet(spaces, 'Pi', sd, r_L)

tnet = tnet * Pi
\end{lstlisting}
To optimize our strategy, we need to simply put it into \texttt{iss\_opt} function:
\begin{lstlisting}[language=python]
qfi, qfis, tnet_opt, status = iss_opt(tnet)
\end{lstlisting}
where \texttt{qfi} is the optimized \acs{QFI}, \texttt{qfis} is a list of pre-QFI values per iteration, \texttt{status} tells whether the optimization converged successfully and \texttt{tnet\_opt} is a copy of \texttt{tnet} where each \texttt{VarTensor} is replaced with a \texttt{ConstTensor} storing the result of the strategy optimization. To access those values we need to use the names of the required tensor and its spaces:
\begin{lstlisting}[language=python]
i = 2

# Accessing the CJ matrix of the i-th
# control operation:
name = f'CONTROL{i}'
# i-th control operation tensor:
C_i = tnet_opt.tensors[name]

spaces = [
    inp[i+1], anc[2*i+1], out[i],
    anc[2*i]
]
# CJ matrix with spaces in the order
# given by the argument:
choi = C_i.choi(spaces)

# Accessing the i-th element
# of the MPS density matrix:
name = rho0_names[i]
# i-th element of the state's
# density MPO:
rho0_i = tnet_opt.tensors[name]

spaces = [
    rho0_bonds[i-1], rho0_spaces[i],
    rho0_bonds[i]
]
# MPS element with indices in the order
# given by the argument:
psi_i = rho0_i.to_mps(spaces)
\end{lstlisting}

Figs.~\ref{fig:passive} (B) and (C) show the optimized values of the \acs{QFI} for collisional strategy together with the parallel and the adaptive ones for comparison. It turns out that in the case of the dephasing noise (B), additional control operations do not help and the results are the same as for the parallel strategy. On the other hand, for the amplitude damping noise (C) the collisional strategy is significantly better than the parallel one and appears to be able to  beat the adaptive strategy for $N>100$, under the given limited ancillary dimensions. This proves that in this case, the optimal strategy found is qualitatively different from both the parallel and the adaptive ones.

\section{Summary}
\label{sec:summary}
We hope that the presented package will serve as an indispensable tool for anyone who wants to identify optimal protocols in noisy metrology scenarios as well as for those who want to compare the performance of strategies of a particular structure with fundamental bounds.

While the package is at the moment focused solely on the single-parameter \acs{QFI} optimization paradigm, it may be in principle developed further to cover Bayesian as well as multi-parameter frameworks. In case of Bayesian framework, though, we do not expect tensor-network methods to work as efficiently as in the \acs{QFI} case. This is mainly due to the need to 
use larger ancilla spaces to effectively feed-forward information that systematically increases with the protocol length, see \cite{kurdzialek2024} for further discussion. 
This would limit applicability of these methods to small-scale problems only, and go against the main purpose of this package, which is to provide numerical methods that work efficiently for large-scale quantum metrology problems. For people interested in small-scale Bayesian estimation problems, we can recommend the QuanEstimation package \cite{Zhang2022}, where implementations of some of the quantum Bayesian estimation methods can be found. 

In general, the QuanEstimation package may be regarded as largely complementary to the present one. It provides a wider set of quantum estimation tools, combining both \acs{QFI}-based and Bayesian methods. The package also has numerous tools dedicated to analyzing and optimizing experimentally relevant quantum control frameworks, as well as some multi-parameter quantum estimation tools. QuanEstimation has optimization procedures that allow to optimize input states in a basic single quantum channel estimation problem, 
as well as optimize adaptive strategies, with some predefined control structures. That said, efficiency of the proposed algorithms is limited to small-scale problems, and even in this case, the optimization procedures employed do not guarantee that the result obtained is indeed optimal.

In case of multi-parameter estimation strategies, one may generalize the  \acs{ISS} and tensor-network methods in a rather straightforward way to optimize the trace of the \acs{QFI} matrix, as in \cite{Albarelli2022}. However, despite some novel works on this topic \cite{Hayashi2024}, a rigorous and at the same time numerically effective optimization of a multi-parameter cost that could be applied to study the performance of noisy metrological protocols in the large $N$ limit seems to be out of reach at this stage.  

Finally, we should also mention that we have independently studied and developed similar methods to study the channel discrimination problem \cite{Bavaresco2021}, results of which will be reported elsewhere.

\emph{Acknowledgments.} We thank Krzysztof Chabuda for fruitful discussions regarding the inner workings of the TNQMetro package. This work has been supported by the ``Quantum Optical Technologies" (FENG.02.01-IP.05-0017/23) project, carried out within the Measure 2.1 International Research Agendas programme of the Foundation for Polish Science co-financed by the European Union under the European Funds for Smart Economy 2021-2027 (FENG), as well as National Science Center (Poland) grant No.2020/37/B/ST2/02134. SK was supported by the Foundation for Polish Science (FNP).

\bibliographystyle{quantum}
\bibliography{bibliography.bib}

\clearpage
\newpage

\appendix

\section{Choi-Jamio{\l}kowski matrix and the link product}\label{sec:link_prod}
When discussing networks of channels it is useful to introduce the notion of the \emph{Choi-Jamio{\l}kowski matrix} (\acs{CJ}). For a linear map
\begin{equation}\label{eq:channel_P}
    \mathrm{P} : \L{\mc A} \mapsto \L{\mc B},
\end{equation}
the corresponding \acs{CJ} matrix is given by \cite{Chiribella2009, Bengtsson2006}:
\begin{equation}\label{eq:choi_P}
    P = \left( \mathrm{P} \otimes \mb{I}_{\mc A} \right)\left(\ket{\Phi}\bra{\Phi}\right) \in \L{\mc B \otimes \mc A},
\end{equation}
where $\ket{\Phi}=\sum_i \ket{i}\otimes \ket{i}$ is a non-normalized maximally entangled state on $\mc A \otimes \mc A$. In what follows, \acs{CJ} matrices of maps (e.g. map $\mr P$) will be denoted with the same symbol but in italics (e.g. $P$). \acs{CJ} matrix $P$ has many useful properties \cite{Bengtsson2006}:
\begin{fact}[\acs{CJ} operator properties]\label{en:choi_properties}
For a map $\mr P$ from \eqref{eq:channel_P} and its \acs{CJ} matrix defined by formula \eqref{eq:choi_P} the following properties are satisfied: 
\begin{enumerate}
    \item if $\mr P$ represents a state preparation procedure then  $P$ is the density matrix of the state it produces,
    \item if $\mr P = \Tr_\mc{A}$ represents the partial trace with respect to $\mc A$ than $P$ is the identity matrix on $\mc A$,
    \item $\mr P$ is completely positive if and only if $P$ is positive semidefinite,
    \item $\mr P$ is trace-preserving if and only if $\Tr_\mc{B}P=\bbone_\mc{A}$,
    \item $\mr P$ is unital, that is $\mr P(\bbone_\mc{A}) = \bbone_\mc{B}$, if and only if $\Tr_\mc{A}P=\bbone_\mc{B}$.
\end{enumerate}        
\end{fact}
\noindent Additionally, for two maps:
\begin{align}\label{eq:chanels_example}
    &\mathrm{P} : \L{\mc H_0} \mapsto \L{\mc H_1 \otimes\mc H_2}, \\
    &\mathrm{Q} : \L{\mc H_2 \otimes \mc H_3} \mapsto \L{\mc H_4},\nonumber
\end{align}
the \acs{CJ} matrix of their composition $\mr R$:
\begin{equation}\label{eq:chanels_example_diag}
    \includegraphics[scale=0.75]{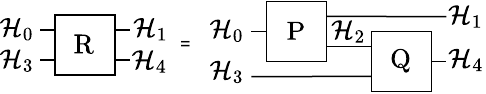},
\end{equation}
is the link product of $P$ and $Q$:
\begin{equation}\label{eq:link_product}
    R = P * Q = \Tr_{2}\left[\left( \bbone_{34} \otimes P^{T_2} \right) \left(Q \otimes \bbone_{01}\right)  \right],
\end{equation}
where $\bbone_{ij}$, $\Tr_{ij}$ and $T_{i}$ denote identity, partial trace and partial transposition on $\mc H_i\otimes\mc H_j$ and $\mc H_i$. In case of \acs{CJ} matrices acting on more complicated Hilbert spaces for example:
\begin{equation}
    M \in \L{\otimes_{i\in X} \mc H_i},\quad N \in \L{\otimes_{i\in Y} \mc H_i},
\end{equation}
where $X$ and $Y$ are some sets of indices, the link product is
\begin{gather}\label{eq:gen_link_product}
    M * N =\\
    =\Tr_{X\cap Y}\left[\left( \bbone_{Y \setminus X} \otimes M^{T_{X \cap Y}} \right) \left(N \otimes \bbone_{X \setminus Y}\right)  \right],\nonumber
\end{gather}
where $\Tr_Z$, $\bbone_Z$ and $T_Z$ are trace, identity matrix and transposition for $\bigotimes_{i\in Z}\mc H_i$ and it is assumed that there are SWAP operators inserted inside the trace such that terms in round brackets act on the tensor product of spaces $\mc H_i$ in the same order. Additionally, since the relation between channels and \acs{CJ} matrices is isomorphic \cite{Chiribella2009, Bengtsson2006} link product can be naturally extended to channels.

\section{Tensor network formalism and tensor representation}\label{sec:tensor_networks}

It is convenient to denote tensors and their contractions using diagrams. In this formalism tensor $T^{\mu_0\mu_1...\mu_{n-1}}$ is represented as a vertex of a graph with an edge for each index:
\begin{equation}
    \includegraphics[scale=0.75]{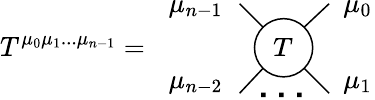},
\end{equation}
and contraction of indices is depicted as a connection of these edges:
\begin{equation}
    \includegraphics[scale=0.75]{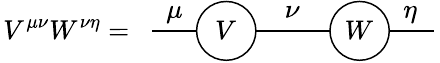}.
\end{equation}

Tensor network formalism can be used to conveniently express the link product.
For a Hilbert space $\mc H=\bigotimes_{i=0}^{n-1}\mc H_i$ with $d_i:=\dim\mc H_i$ and a linear operator acting on it:
\begin{equation}\label{eq:matrix_tensor}
    M = \sum_{\substack{a_0, ..., a_{n-1}\\b_0, ..., b_{n-1}}} M^{a_0 ... a_{n-1}}_{b_0 ... b_{n-1}} \ket{a_0 ... a_{n-1}}\bra{b_0 ... b_{n-1}},
\end{equation}
we define the \emph{tensor representation} of $M$ to be a tensor:
\begin{equation}
    \tilde M^{x_0...x_{n-1}} = M^{a_0 ... a_{n-1}}_{b_0 ... b_{n-1}} \quad \mr{for} \quad x_i=d_ia_i+b_i.
\end{equation}
Next, consider channels $\mr P$, $\mr Q$ and $\mr R = \mr P * \mr Q$ from \eqref{eq:chanels_example} and \eqref{eq:chanels_example_diag}. One can easily check that tensor representations of their \acs{CJ} matrices satisfy:
\begin{equation}
    \includegraphics[scale=0.75]{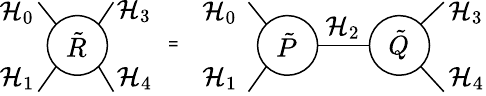},
\end{equation}
which mirrors the circuit diagram from \eqref{eq:chanels_example_diag}. Due to the similarity of these two diagrams, in the main text we drop the symbol $\sim$ and we depict strategies using circuit diagrams and tensor network diagrams interchangeably.

When $M$ is an \acs{MPO}:
\begin{equation}\label{eq:mpo}
    \includegraphics[scale=0.75]{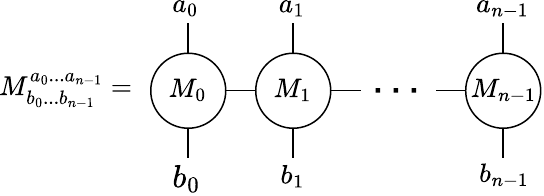}.
\end{equation}
we can extend this definition to its components $M_i$ such that $\tilde M$ will be a result of contraction of tensors $\tilde M_i$:
\begin{gather}
    \includegraphics[scale=0.75]{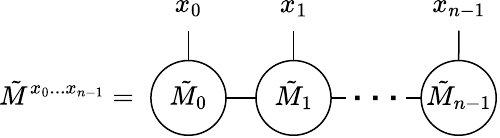},\\
    \includegraphics[scale=0.75]{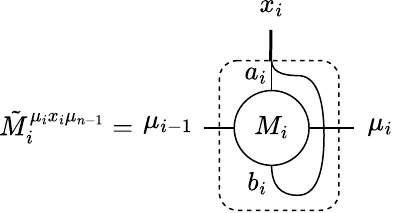},
\end{gather}
that is, only physical indices of $M_i$ are joined. This, combined with the use of `*' to denote both link product and tensor contraction, allows for convenient depiction of link product with \acs{CJ} matrices which are \acs{MPO}s.


\section*{List of acronyms}

\begin{acronym}
    \acro{AD}{Adaptive strategy}
    \acro{BFS}{Breadth First Search}
    \acro{CCR}{Classical Cram{\'e}r-Rao}
    \acro{CFI}{Classical Fisher Information}
    \acro{CJ}{Choi-Jamio{\l}kowski}
    \acro{CPTP}{Completely Positive Trace-Preserving}
    \acro{DAG}{Directed Acyclic Graph}
    \acro{FIFO}{First-In First-Out}
    \acro{FOM}{Figure of Merit}
    \acro{HS}{Heisenberg Scaling}
    \acro{ISS}{Iterative See-Saw}
    \acro{LFI}{Local Fisher Information}
    \acro{MOP}{Minimization Over Purifications}
    \acro{MPO}{Matrix Product Operator}
    \acro{MPS}{Matrix Product State}
    \acro{PAR}{Parallel Strategy}
    \acro{PAD}{Passive Adaptive Strategy}
    \acro{PC}{Personal Computer}
    \acro{QCR}{Quantum Cram{\'e}r-Rao bound}
    \acro{SDP}{Semidefinite programming}
    \acro{SR}{Simple Repetitions strategy}
    \acro{SS}{Standard Scaling}
    \acro{SLD}{Symmetric Logarithmic Derivative}
    \acro{QFI}{Quantum Fisher Information}
    \acro{UML}{Unified Modeling Language}
\end{acronym}

\end{document}